\documentstyle[12pt, epsf]{article}

\begin{document}
\baselineskip=0.7cm
\newcommand{\EQ}{\begin{equation}}
\newcommand{\EN}{\end{equation}}
\newcommand{\EQA}{\begin{eqnarray}}
\newcommand{\EQN}{\end{eqnarray}}
\newcommand{\e}{{\rm e}}
\newcommand{\Sp}{{\rm Sp}}
\renewcommand{\theequation}{\arabic{section}.\arabic{equation}}
\newcommand{\Tr}{{\rm Tr}}
\newcommand{\lpartial}{\buildrel \leftarrow \over \partial}
\newcommand{\rpartial}{\buildrel \rightarrow \over 
\partial}

\renewcommand{\thesection}{\arabic{section}.}
\renewcommand{\thesubsection}{\arabic{section}.\arabic{subsection}}
\makeatletter
\def\section{\@startsection{section}{1}{\z@}{-3.5ex plus -1ex minus
 -.2ex}{2.3ex plus .2ex}{\large}}
\def\subsection{\@startsection{subsection}{2}{\z@}{-3.25ex plus -1ex minus
 -.2ex}{1.5ex plus .2ex}{\normalsize\it}}
\def\appendix{
\par
\setcounter{section}{0}
\setcounter{subsection}{0}
\def\thesection{\Alph{section}}}
\makeatother
\def\thefootnote{\fnsymbol{footnote}}
\begin{flushright}
hep-th/0108176\\
UT-KOMABA/01-04\\
TIT/HEP-469\\
August 2001
\end{flushright}
\vspace{1cm}
\begin{center}
\Large
From Supermembrane to Matrix String
 
\vspace{1cm}
\normalsize
{\sc Yasuhiro Sekino}
\footnote{
e-mail address:\ \ {\tt sekino@th.phys.titech.ac.jp}}
\vspace{0.3cm}

{\it Department of Physics, Tokyo Institute of Technology\\
Ookayama, Meguro-ku, Tokyo 152-8550}

\vspace{0.4cm}
and 

\vspace{0.4cm}
{\sc Tamiaki Yoneya}
\footnote{
e-mail address:\ \ {\tt tam@hep1.c.u-tokyo.ac.jp}}
\\
\vspace{0.3cm}

{\it Institute of Physics, University of Tokyo\\
Komaba, Meguro-ku, Tokyo 153-8902}

\vspace{1cm}
Abstract
\end{center}
We develop a
 systematic method 
of directly embedding supermembrane wrapped around a circle 
into matrix string theory. 
Our purpose is to study 
connection between matrix string and membrane from an entirely 11 dimensional point of view. The method 
does neither rely upon the DLCQ limit nor 
upon string dualities.  In principle, 
this enables us to construct matrix string theory 
with arbitrary backgrounds from the 
corresponding supermembrane theory. 
As a simplest application 
of the formalism, the matrix-string action 
with a 7 brane background (Kaluza-Klein Melvin solution) 
with nontrivial RR vector field is given. 

\newpage
\section{Introduction}
Supermembrane is expected to play  a pivotal role 
in the quest for the fundamental degrees of freedom 
of the conjectured M-theory. For example, 
in Matrix theory \cite{bfss} as the first 
concrete proposal along such direction, 
the matrix-regularized form of 
light-cone supermembrane action 
is reinterpreted as the effective theory 
of D-particles in the infinite-momentum 
frame boosted along the compactified 
11th direction. In particular, the 
diagonal elements of 
matrix coordinates are identified with the transverse 
coordinates of D-particles. 
This means that the 
membrane itself  is boosted along the 
compactified direction as a whole. 
In this picture, it is not possible to see 
fundamental strings directly. Although 
we assume that the off-diagonal elements 
would correspond to fundamental open strings attached 
to D-particles, it is difficult to 
exhibit such properties from the viewpoint of 
the regularization of membrane world volume. 
In the context of the so-called DLCQ limit, 
this is not unreasonable, 
since in this limit the length of 
open strings connecting D-particles 
must be regarded as being far too shorter than the 
typical string scale $\ell_s$, and therefore the 
stringy behavior of the open strings does not 
manifest itself. In spite of some nontrivial confirmations to 
two-loop order \cite{bbpt}\cite{okayone} 
on the behaviors of graviton in the DLCQ limit, 
it is not at all clear 
whether this theory gives the complete description of 
gravity in 11 (and even in 10) dimensional space-time. 

On the other hand, in matrix {\it string} theory \cite{dvv-motl} which 
followed the proposal of Matrix theory, 
fundamental strings regain the role of 
the basic degrees of freedom. This was originally 
explained, on the  basis of the Matrix theory 
conjecture, by combining T and S-dualities
with the  flipping of the compactified direction 
from 11th to 9th, which is now transverse to the 
light-like directions. Namely, the T-duality 
converts D-particles into D-strings. Then, 
D-strings are 
 turned into fundamental (F) strings  
 after using S-duality and making an inverse 
T-duality transformation. Also the 
D-particle quantum number is understood as electric flux. 
Direct visibility 
of fundamental strings in this way ensures that 
gravity is consistently described in the 
sense of 10 dimensional space-time, to the extent that  matrix string 
is reducible to ordinary light-cone string \cite{wynteretal} in 
the weak string coupling limit $g_s \rightarrow 0$. 
However, as in the case of Matrix theory, 
the situation in 11 dimensions is quite obscure. 

Now if we recall that fundamental string can 
also be understood, at least classically, 
 as double-dimensionally 
reduced membrane \cite{dhis} from 11 dimensions, 
it is natural to ask whether and 
how precisely matrix string is related to such reduced 
membrane. We expect that the wrapped membrane 
should be directly mapped to matrix string 
{\it without} invoking duality arguments. 
This is not a trivial question, and to the best of our knowledge, 
 such a possibility has never 
been studied in the published literature. 
We think that clarification of this 
correspondence  will be important and useful for at least 
three reasons:
(1)  First of all, it is of intrinsic interest of its own to check 
overall consistency between the web of 
string dualities 
and the membrane-string connection from a purely 
11 dimensional viewpoint. 
(2)  It would provide a new hint on the treatment of 
the dynamics both of membrane and of 
matrix string, especially, with respect to the nature 
of the large $N$ limit of matrix string theory, and 
also on the nature of its decompactification limit 
$g_s\rightarrow \infty$, by clarifying the meaning and 
role of the off-diagonal elements of 
matrix string variables. 
(3)  From a more practical point of view, it would 
help us to formulate the 
matrix string theory on general backgrounds,  
given the corresponding formulation of 
supermembrane, going beyond linearized 
approximation. All of these can be a first step 
towards  the more crucial task of deriving 
11 dimensional gravity from membrane-matrix-string 
theories.

With this motivation, we develop a systematic method of directly relating
 the wrapped (super)membrane to matrix string. 
Basically, we show that the off-diagonal matrix 
variables of matrix string theory are directly 
identified, in the large $N$ limit,  
with the higher Kaluza-Klein modes with 
respect to the world-volume momentum along the 
wrapped direction. 
This provides us a general prescription 
of embedding arbitrary membrane 
action into matrix-string action 
in the large $N$ limit. 
We also briefly study the nature of the 
double dimensional reduction of supermembrane
 quantum-mechanically. 
Our discussion shows that the 
double-dimensional reduction is a subtle 
problem which is common to the 
infra-red reduction from matrix string to 
the ordinary light-cone string theory in the 
weak coupling limit. 
As a simplest application of the general  correspondence 
for the extension of the theory to nontrivial backgrounds, 
we discuss the matrix-string action 
in the background, so-called Kaluza-Klein Melvin solution, representing a 7-brane
with a Ramond-Ramond  (magnetic) vector 
field of arbitrary strength. 

In the next section, we start from a brief review of 
light-cone supermembrane theory and then 
discuss its compactification on a circle and 
quantum mechanical reduction. Some preliminary 
discussions on the quantum-mechanical 
double dimensional reduction, which is based on a 
strong-coupling expansion, is given  
 in Appendix A. 
In section 3, we discuss the correspondence of membrane 
with the matrix string and present  a general 
formula which enables us to map arbitrary 
trace-integrals of the matrix string variables into the 
corresponding integrals over the 
membrane volume.  
In section 4, we derive the matrix-string action 
for a 7-brane background with magnetic RR vector field. 
The relation of our result with previous 
works in the linearized approximation is briefly summarized 
in Appendix B. In the  final section, 
we conclude the paper with discussions  
on remaining problems and  future possibilities. 

\vspace{0.3cm}
\section{Light-cone supermembrane and its 
compactification}
  
It is well known that the light-cone dynamics \cite{dwhn} 
of supermembrane,  
 in the sector of fixed total longitudinal momentum $P^+$, 
is summarized by
the  following effective action. 
\[
A = {1\over \ell_M^3}\int d\tau \int_0^{2\pi L}d\sigma \int_0^{2\pi L} d\rho \, \, {\cal L}, 
\]
\EQ
w^{-1}{\cal L}
={1\over 2}(D_0 X^a)^2 +i \overline{\psi}\gamma_-D_0 \psi
-{1\over 4}(\{X^a, X^b\})^2 
+i\overline{\psi}\gamma_-\Gamma_a\{X^a, \psi \} ,
\label{lightconemembrane}
\EN
where the covariant time derivative $D_0$ is 
with respect to gauge field $A_0$ :
$D_0 X^a= \partial_{\tau} X^a-\{A_0, X^a\}$ 
and the spatial index $a$ runs through $1,2,\ldots, 9$.  
The density function $w$ is introduced in the 
gauge fixing process such that the 
longitudinal momentum $P^+(\sigma, \rho)$ 
satisfies $P^+(\sigma, \rho)=P^+w(\sigma, \rho)/L^2$. 
The  bracket notation is defined by 
\[
\{X^a, X^b\}=w^{-1}(\partial_{\sigma}X^a\partial_{\rho}
X^b -\partial_{\rho}X^a\partial_{\sigma} 
X^b) .
\]
The time coordinate 
$\tau$ is related to the light-cone time by 
\EQ
\ell_M^3P^+\tau/(2\pi L)^2=X^+, 
\label{timerelation}
\EN
 such 
that the total center of mass (transverse) momentum $P^a$ 
contributes to the Hamiltonian 
in the standard form, 
\[
P^-dX^+ =\Big({(P^a)^2\over 2P^+} + \cdots\Big)dX^+.
\] 
The membrane tension $1/\ell_M^3$ is assumed to be 
the fundamental M-theory scale, namely  
$\ell_M=g_s^{1/3}\ell_s$ in terms of the standard 
string theory parameters up to some numerical 
constant. We assumed that the 
space-time dimension is 11. 
This is  justified if we can 
establish the validity of the double-dimensional 
reduction quantum-mechanically in the limit of 
small compactification radius, as we discuss later. 
We use the purely real Majorana representation 
for the $\Gamma$ matrices 
($\gamma_{\pm}=(\Gamma^{10}\pm 
\Gamma^0)/\sqrt{2},  (\Gamma^0)^T=
-\Gamma^0, (\Gamma^{10})^T=\Gamma^{10}, (\Gamma^a)^T
=\Gamma^a$). 
In ref. \cite{dwhn}, everything is
dimensionless.  We recovered dimensions by 
normalizing the  spatial world-volume parameters $(\sigma,
\rho)$ as 
\[
0 \le \sigma \le 2\pi L, \quad 0\le \rho\le  2\pi L ,
\quad 
\int d\sigma d\rho\, \,  w=L^2
\]
with $L$ being some arbitrary length parameter, which will  
later be chosen to be  the radius of the compactification circle. 
Note that the arbitrariness of $L$ is manifest in the action: 
$L$ can be
eliminated from the action  by performing the rescaling $
\tau\rightarrow L^2\tau, \sigma \rightarrow L\sigma, 
\rho\rightarrow L\rho, \psi \rightarrow \psi/L$. 
For simplicity, we choose the density function $w(\sigma, 
\rho)$ to be 
constant so that $w=(2\pi)^{-2}$,  by assuming that the 
topology of membrane is simply torus. In this
convention,  the Lagrangian density is dimensionless. 
The supersymmetry transformation law is given by 
\EQA
\delta_{\epsilon} X^a &=&-2i\overline{\epsilon}\Gamma^a \psi , \nonumber \\
\delta_{\epsilon} \psi &=& {1\over 2}\gamma_+
(D_0 X^a \Gamma_a +\gamma_-)\epsilon
+{1\over 4}
\{X^a, X^b\}\gamma_+\Gamma_{ab}\epsilon ,
\label{susylaw}
\\
\delta_{\epsilon} A &=& -i2\overline{\epsilon}\psi .
\nonumber
\EQN

The Gauss-law constraint derived from  
this action by the variation 
with respect to the gauge field $A_0$ gives the 
constraint corresponding to the 
area-preserving diffeomorphism (APD) 
which is the residual reparametrization symmetry
$\delta X^a =\{\Lambda, X\}, \, \, \delta 
A_0 =\partial_0 \Lambda+\{\Lambda, A_0\}, \, etc$, 
 after 
fixing to the light-cone gauge. More precisely, 
the Gauss-law constraint
\EQ
\{D_0 X^a, X^a\} +i\{\overline{\psi}, \gamma_-\psi\}=0
\label{gausslaw}
\EN
is the integrability condition for the equation 
determining the longitudinal coordinate $X^-$.  
The latter is 
\EQ
 {\ell_M^3\over (2\pi L)^2}P^+\partial_jX^-
+\partial_{\tau}X^a \partial_j X^a 
+i\overline{\psi}\gamma_-\partial_j\psi=0
\EN
which is locally equivalent with the condition 
(\ref{gausslaw}).  When there exist topologically 
nontrivial cycles on membrane spatial world-volume, 
we have to further impose the 
global constraint of the form
\EQ
\oint d{\sigma}^j \, (\partial_{\tau}X^a \partial_j X^a 
+i\overline{\psi}\gamma_-\partial_j\psi) =0
\label{cyclecondition}
\EN
to ensure that $X^-$ is periodic along the cycles.\footnote{
In ref. \cite{dwpp}, it was emphasized that we should 
supplement the term $D_0X^-$  to the lagrangian in the 
presence of winding. We drop 
this term, since it is a total derivative, as long as we 
assume periodicity for $X^-$,  and hence does not play 
any substantial role below.  }
Of course, if the $X^-$ itself is compactified 
as in the DLCQ treatment which is {\it not} adopted 
in the present paper, the right-hand side should
be  proportional to integers times the  corresponding compactification 
radius.  

Now we study the situation where one, 9th, of the transverse 
directions is compactified along a circle by 
assuming this direction to be the `eleventh' direction 
of M-theory. 
We set the radius  of the circle to be 
$L=g_s\ell_s$, by identifying it with the arbitrary 
parameter of the membrane action. Then, of course, 
$L$ becomes a physical parameter.  
Denoting the 9-th coordinate by 
$Y=X^9(\sigma, \rho, \tau)$ and choosing the  world-volume 
coordinate $\rho$ along the $Y$-direction, 
the compactification amounts to the  condition
\EQ
Y(\sigma, \rho+2\pi L, \tau)=2\pi L + Y(\sigma, \rho, \tau)
\label{compcondition} .
\EN
We denote the remaining eight transverse directions 
$(1, 2, \ldots, 8)$ by the indices $i, j, k, \ldots.$

Classically, the double dimensional reduction 
assumes that everything is $\rho$-independent, 
\[
\partial_{\rho}X^i =0, \quad \partial_{\rho}A=0, \quad
\partial_{\rho}\psi =0, \quad
\partial_{\rho}Y=1.
\]
The action then reduces to 
\EQ
A={2\pi L\over \ell_M^3}\int d\tau \int_0^{2\pi L}
 d\sigma \Big[
{1\over 2}(\partial_0 X^i)^2 -
{1\over 2}(2\pi)^4(\partial_{\sigma} X^i)^2 
+i\overline{\psi}\gamma_-\partial_0\psi
-i(2\pi)^2\overline{\psi}\gamma_-\Gamma_9\partial_{\sigma}\psi
\Big] .
\EN
By the identification ${(2\pi)^3 L/\ell_M^3}=1/2\pi
\alpha'$ which is kept finite in the limit $L\rightarrow 0$, and by rescaling $\sigma \rightarrow L\sigma, 
\tau\rightarrow L\tau/(2\pi)^2$ and also by a change of
normalization 
of the fermionic coordinate $\psi$,  
 this is nothing but
the standard world-sheet action of  the light-cone superstring 
in the Green-Schwarz formalism. 

The naive double-dimensional reduction completely
ignores  the nonzero momenta along the 
compactified direction $\rho$. 
It is, however,  important to note that 
quantum mechanically the suppression of higher 
momentum (Kaluza-Klein) modes along the $\rho$ direction 
can not be so straightforward.  
One might naively expect that, just as the usual 
Kaluza-Klein compactification in ordinary 
field theory, 
the  modes with nonzero momenta along the 
compactified direction would be 
suppressed by the large masses of order 
$1/L$ viewed  from lower dimensional space. 
This is not the case in the double 
compactification of membrane. It is simply not  
possible to apply this standard argument to 
the membrane, since there arises no mass term 
in the usual sense. Instead of the ordinary 
mass term, we  have the leading quadratic terms 
in the potential part of the action as 
\[
{1\over 4}\{X^i, X^j\}^2
\rightarrow {1\over 4}\Big((\partial_{\sigma}x^i +
\partial_{\sigma}X^i) 
\partial_{\rho}X^j 
-(\partial_{\sigma}x^j +
\partial_{\sigma}X^j) 
\partial_{\rho}X^i 
\Big)^2
\]
\EQ
\rightarrow {1\over 2}(\partial_{\sigma}x^i)^2
(\partial_{\rho}X^j)^2-
{1\over 2}(\partial_{\sigma}x^i \partial_{\rho}X^i)^2 + \cdots , 
\label{leadingterm}
\EN
where we have separated the zero mode part $x^i(\sigma)$ 
by making the shift $X^i(\sigma) \rightarrow x^i(\sigma) 
+X^i(\sigma, \rho)$. Naively, this form behaves 
as $\sigma$-dependent mass terms for nonzero modes 
$X^i$. 
However, whether ignoring higher modes on the 
basis of this is justified seems a subtle
question.\footnote{
The subtlety here is of the same nature 
as has been discussed in 
ref. \cite{hms} for  infrared reduction 
of $D=4, N=4$ Yang-Mills theory to 
a conformal field theory. The problem 
is common to the reduction to orbifold 
string from matrix string. 
}  A possible approach is to show explicitly that the effective
theory  after integrating  over the infinite set of the 
higher modes is given by the superstring action in 
the limit $L\rightarrow 0$.  
Since $i\partial_{\rho} \sim O(1/L)$, we see that 
the strength of the fluctuations
 of nonzero Kaluza-Klein  modes 
is in fact proportional  $O(L)$. However, the Gaussian 
fluctuation generically gives rise to an order $O(L^0)$ 
contribution to the action, owing to the 
dependence of the coefficients on 
the ($\sigma$-dependent) zero modes. This cannot 
be neglected since the reduced action 
itself is of the same order $O(L^0)$ as long 
as $\alpha'$ is fixed in the limit $L\rightarrow 0$.  

Note that the relevant expansion 
here is essentially the {\it strong-coupling} expansion 
with respect to the (gauge) coupling $g\sim 1/L
=1/g_s\ell_s \rightarrow \infty$ as $g_s \rightarrow 0$. 
The weak coupling perturbation theory 
(or semi-classical argument) is not suitable. 
As discussed in Appendix A, 
it is easy to check that, in the leading order in 
strong-coupling $1/g\sim L$ expansion, 
integrations over the 
higher modes precisely cancel between the 
bosonic and fermionic modes, leaving the 
double-dimensionally reduced 
action, except possibly at the point where $(\partial_{\sigma}x^i)^2$ 
vanishes. Such points would just correspond to where 
 interactions are occuring by the 
joining or splitting of strings. 
In the membrane picture with one 
more dimension, this kind of topology change 
would correspond to 
a smooth dynamical process at the vicinity of 
the interaction point. 

However, strong coupling perturbative expansions in continuum 
field theory such as we encounter here are necessarily singular due to its
ultra-local character,   and hence it is not easy to 
rigorously justify the double-dimensional 
reduction in this way. We give a preliminary 
discussion on the higher-order effects in 
the strong coupling expansion in Appendix A. 
In any case, our argument implies that supersymmetry is very crucial 
for justifying the double-dimensional reduction 
quantum-mechanically: 
In the limit of vanishing string coupling 
$L\rightarrow 0$, 
the susy transformation law for the zero modes reduces to the 
ordinary linear transformation law,  
since the nonlinear term $\{X^a, X^b \}\Gamma_{ab}\epsilon$ 
which is the only possible term correcting the 
susy transformation law 
is of order $L^2/L =L$, and hence the 
zero mode action must take the ordinary 
form without any correction. This also shows that the 
critical dimension of supermembrane theory 
is  $D=11$ in the limit of 
vanishing compactification radius.\footnote{
In preparing the present manuscript, the work  \cite{ishihaya} 
appeared, discussing the critical 
dimension of bosonic membrane on the basis of the 
ordinary  perturbative treatment of (bosonic) membrane. }
 In the case of bosonic membrane, there is no 
cancellation as above and the integration 
gives complicated (singular) contribution for the 
zero mode $x^i$, and hence it does not seem 
possible to justify the double-dimensional reduction 
quantum-mechanically. 

Justification of the double-dimesional reduction 
is not the main concern of the present work. Rather, 
we would like to establish some direct relation between the compactified 
membrane and the 
large $N$ limit of matrix string theory.  
To prepare for our discussion about this direction in the 
next section,  
let us make more explicit how the  compactified 
direction $Y$   
behaves in the action. Using the same scaled variables as the 
naive double-dimension, 
the full action is 
\EQA
A={(2\pi)^2/\ell_M^3}&&\hspace{-0.7cm}
\int d\tau \int_0^{2\pi}
d\sigma
\int_0^{2\pi L} d\rho \Big(
{1\over 2}(D_0X^i)^2+{1\over 2}(D_0Y)^2 
-{1\over 4}\{X^i, X^j\}^2
-{1\over 2}\{X^i, Y\}^2 \nonumber \\
&&+i\psi^T D_0\psi
+i\psi^T\Gamma_i\{X^i,\psi\}
+i\psi^T\Gamma_9\{Y, \psi\}
\Big) ,
\label{Yaction}
\EQN
where the Poisson bracket is now defined as 
$\{A, B\}=\partial_{\sigma}A\partial_{\rho}B-
\partial_{\rho}A\partial_{\sigma}B$ because of the 
rescaling. Note that in this form the action 
is invariant with respect to a 
global rescaling of two-dimensional space $(\tau, \sigma)
\rightarrow (\lambda\tau, \lambda\sigma)$ with $ 
\psi \rightarrow \sqrt{\lambda}^{-1}\psi$. 
The light-cone 
gauge condition $\gamma_+ \psi=0$ implies 
$\overline{\psi}\gamma_-=
\sqrt{2}\psi^T$. We have changed the 
normalization of the fermion field 
to eliminate the factor $\sqrt{2}$.  

To take  the condition of compactification into account, 
we redefine the $Y$ coordinate by making a shift 
\EQ
Y\rightarrow \rho +Y .
\EN 
After this shift, all world-volume fields are 
assumed to be periodic with respect to $\rho$. 
The terms affected by this substitution are 
\[
{1\over 2}(D_0 Y)^2\rightarrow {1\over 2}(\partial_0 Y-
\partial_{\sigma}A -
\{A, Y\})^2  ,
\]
\[
{1\over 2}\{X^i, Y\}^2 \rightarrow {1\over
2}
(\partial_{\sigma} X^i -\{Y, X^i\})^2  ,
\]
\[
\psi^T\Gamma_9\{Y, \psi\}
\rightarrow -\psi^T\Gamma_9(\partial_{\sigma}\psi 
-\{Y, \psi\}) .
\]
 Thus the final form of the action 
is, after rescaling $\rho$ by $\rho 
\rightarrow L\rho$, 
\EQA
A={(2\pi)^2L/\ell_M^3}&&\hspace{-0.7cm}
\int d\tau \int_0^{2\pi}
d\sigma
\int_0^{2\pi} d\rho \nonumber \\
&&\Big[
{1\over 2}F_{0, \sigma}^2+
{1\over 2}(D_0X^i)^2 -{1\over 2}(D^Y_{\sigma}X^i)^2
-{1\over 4L^2}\{X^i, X^j\}^2
  \nonumber \\
&&+i\psi^T D_0\psi -i\psi^T\Gamma_9D_{\sigma}^Y\psi
+i{1\over L}\psi^T\Gamma_i\{X^i,\psi\}
\Big]
\label{doubledimmemaction}
\EQN
where 
now the covariant derivatives and the field strength are 
defined as 
\EQ
F_{0, \sigma}=\partial_0 Y- \partial_{\sigma}A-
{1\over L}\{A, Y\},
\EN
\EQ
D_0X^i =\partial_0  X^i  -{1\over L}\{A, X^i\} ,
\EN
\EQ
D^Y_{\sigma}X^i =\partial_{\sigma}X^i -
{1\over L}\{Y, X^i\}
\EN
and similarly for $\psi$'s. This is nothing but 
the two-dimensional gauge theory of APD, 
where $(A_0=A, A_1=A_{\sigma}= Y)$ plays the role of gauge field 
and the inverse compactification radius $1/L$ is the 
gauge coupling constant. The (infinitesimal) gauge transformation 
is 
\EQ
\delta A_r= L\partial_r \Lambda +\{\Lambda, A_r\},\ \, 
\delta X^i= \{\Lambda, X^i\},\, \, 
\delta \psi= \{\Lambda, \psi\}.
\label{gaugetrsf}
\EN

At this point, the reader must recognize that 
the structure exhibited here is very close to
 that of matrix string theory. 
Indeed, if Poisson bracket is replaced 
by  matrix commutator, it seems that the above action 
formally reduces to the matrix string action. 
However, the usual correspondence 
between U($N$) matrices and the two-dimensional 
phase space $(\sigma, \rho)$ does not work here. If it worked, 
the theory would have been reduced to $0+1$ dimensional 
matirx theory, but matrix string theory is a two(=1+1)-dimensional 
gauge theory.  A resolution of this small puzzle will be 
given in the next section by establishing  a new direct 
correspondence between the two, 
which is an extension 
of the familiar correspondence 
introduced in \cite{dwhn}.

\vspace{0.4cm}
\section{Correspondence of wrapped membrane with matrix string}
\setcounter{equation}{0}
In order to motivate our method, let us first 
start from considering the
Poisson bracket between a zero ($x(\sigma)$) 
and a nonzero mode part $X(\sigma, \rho)$ 
which is Fourier-decomposed as 
\[
X^j(\sigma, \rho)= 
\sum_n X_n^j(\sigma)\e^{in\rho} ,  
\]
\EQ
\{x^i, X^j\}(\sigma, \rho) =\partial_{\sigma}x^i
\partial_{\rho}X^j(\sigma,\rho)
 =\sum_n \partial_{\sigma}x^i(\sigma) X_n^j(\sigma)
in e^{in\rho}  .
\label{diaoffdiabracket}
\EN
We compare this expression with the commutator 
in matrix string theory:
\EQ
[x^i(\theta), X^j(\theta)]_{nm}=(x^i_n(\theta) -
x^i_m(\theta))X^j_{nm}(\theta)
\label{matstxXcommutator}
\EN
between $x^i$ with only diagonal matrix
elements 
$x^i_{nn}\equiv x^i_n$ and a generic matrix with 
nonzero off-diagonal elements. 

Suppose we
consider a long  string which satisfies the orbifold
condition 
\EQ
x^i_k(\theta +2\pi) =x^i_{k+1}(\theta), \quad (k=
1, 2, \ldots, N-1),  \quad 
x^i_N(\theta+2\pi)=x_1^i(\theta)
\EN
for $N\gg 1$. Then it is natural to identify the diagonal 
components $x^i_n(\theta)$ with the membrane zero mode 
$x^i(\sigma)$ by 
\EQ
x^i(\sigma) = \left\{
\begin{array}{ll}
x^i_1(\theta) , & 0\le \sigma \le 2\pi/N \\
x^i_2(\theta) , & 2\pi/N\le \sigma \le 4\pi/N \\
\quad . & \\
\quad . & \\
\quad . & \\
x^i_N(\theta) , & 2(N-1)\pi/N \le \sigma \le 2\pi  .
\end{array}
\right.
\label{decomposition}
\EN
For sufficiently large $N$ and for generic $k, \ell$ such that 
$|k-\ell| \ll N$,  
this leads to 
\EQA
x^i_k(\theta)-x^i_{\ell}(\theta) 
&=&x^i(\sigma_{k\ell}+{1\over 2}(\sigma_k-\sigma_{\ell}))
-x^i(\sigma_{k\ell}-{1\over 2}(\sigma_k-\sigma_{\ell}))
\nonumber
\\
&=&(\sigma_k-\sigma_{\ell})\partial_{\sigma}x^i
(\sigma_{k\ell})
={2(k-\ell)\pi\over N}\partial_{\sigma}x^i(\sigma_{k\ell}) ,
\EQN
where we set as 
\EQ
\sigma_k={2(k-1)\over N}\pi +{\theta \over N} , 
\quad 
\sigma_{\ell}={2(\ell-1)\over N}\pi +{\theta \over N}
\label{segment} 
\EN
and 
\EQ
\sigma_{k\ell}=(\sigma_k + \sigma_{\ell})/2 .
\EN
 This shows that the commutator (\ref{matstxXcommutator}) 
of matrix-string theory and the Poisson bracket of 
doubly compactified membrane 
is identical in the large $N$ limit under the 
correspondence
\EQ
 \{x^i, X^j\}_n (\sigma) \equiv 
\int {d\rho\over 2\pi}  e^{-in\rho}\{x^i, X^j\} 
(\theta, \rho)
\leftrightarrow  
i({2\pi \over N})^{-1}[x^i, X^j]_{k\ell}(\theta)
\EN
with  
\EQ
n=k-\ell, \quad \sigma = \sigma_{k\ell}={k+\ell -2\over
N}\pi  +{\theta\over N} ,
\label{nsigmathetacorres}
\EN
by identifying the matrix element $X_{k\ell}$ 
with the $n (=k-\ell)$-th Fourier mode 
\EQ
X^j_n(\sigma)
=X^j_{k\ell}(\theta), 
\EN
which obeys the condition 
\EQ
X^i_{k\ell}(\theta+2\pi) =X^i_{k+1\, \ell+1}(\theta)
\label{bcond}
\EN 
corresponding to the decomposition (\ref{decomposition}). 
This provides us a first hint for a direct mapping between 
membrane and matrix string. Namely, we start from the 
the diagonal elements (zero modes) and include 
the off-diagonal elements 
(higher Kaluza-Klein (KK) Fourier 
modes), gradually from near to far off-diagonals.  

Our next task is to check whether this correspondence can 
be generalized to brackets between arbitrary 
functions of $\rho$.  Let us first rewrite the general 
commutator between two matrices 
with nonzero off-diagonal components as 
\EQA
[X^i, X^j]_{k\ell}&&\hspace{-0.5cm}
=X^i_{km}X^j_{m\ell}-X^j_{km}X^i_{m\ell}\nonumber \\
&&\hspace{-0.5cm}
=X^i_{k-m}(\sigma_{km})X^j_{m-\ell}(\sigma_{m\ell})
- X^j_{k-m}(\sigma_{km})X^i_{m-\ell}(\sigma_{m\ell}) 
\EQN
Then, by using 
\[
\sigma_{km}=\sigma_{k\ell}+{m-\ell\over N}\pi, 
\quad 
\sigma_{m\ell}=\sigma_{k\ell}+{m-k \over N}\pi , 
\]
we find ($\sigma = \sigma_{k\ell}$)
\EQA
[X^i, X^j]_{k\ell}(\theta) &&\hspace{-0.5cm}
=X^i_{km}(\theta)X^j_{m\ell}(\theta)-
X^j_{km}(\theta)X^i_{m\ell}(\theta)\nonumber
\\ &&\hspace{-0.5cm}
=X^i_{k-m}(\sigma)X^j_{m-\ell}(\sigma)-
X^j_{k-m}(\sigma)X^i_{m-\ell}(\sigma) \nonumber \\
&&
\hspace{-1.5cm}+X^i_{k-m}(\sigma){(m-k)\pi\over
N}\partial_{\sigma}X^j_{m-\ell} (\sigma)
+{(m-\ell)\pi\over N}\partial_{\sigma}X^i_{k-m}
(\sigma) X^j_{m-\ell}(\sigma)
\nonumber \\
&&
\hspace{-1.5cm}
-X^j_{k-m}{(m-k)\pi\over N}\partial_{\sigma}
X^i_{m-\ell}(\sigma)-{(m-\ell)\pi\over N}
\partial_{\sigma}X^j_{k-m}(\sigma) X^i_{m-\ell}(\sigma) 
\\
&& + O(N^{-2}) . \nonumber
\EQN
After the summation over $m$ 
(making the shift $m\rightarrow -m +\ell +k$), the first  line 
vanishes, while the second and the third terms 
give the $n (=k-\ell)$-th Fourier mode of the 
corresponding Poisson bracket, 
\EQ
i({2\pi\over N})^{-1}[X^i, X^j]_{k\ell} 
\Leftrightarrow 
\int {d\rho\over 2\pi}e^{-in\rho}
\{X^i, X^j\}
\label{commpoisson}
\EN
with the same identification (\ref{nsigmathetacorres}) 
between the mode numbers in the 
large $N$ limit and the world-volume coordinates.  

The dictionary of the correspondence  is thus summarized as 
\vspace{0.5cm}
\begin{center}
\begin{tabular}{c|c}
 Long string of matrix string theory &  
Doubly compactified membrane \\ \hline\hline
& \\

$\Tr {1\over N}\int^{2\pi}_0 d\theta $ & 
$\int_0^{2\pi} d\sigma {1\over 2\pi}\int_0^{2\pi} d\rho$ \\ 

& \\
$i{N\over 2\pi}[A, B]$ & $\{A, B \}$ \\

& \\ \hline

& \\

$A_{k\ell}(\theta)$ & 
$\int {d\rho\over 2\pi}  e^{-in\rho} A(\sigma, \rho)  $ \\

& \\
$k-\ell$ & $n$ \\

&\\
$\theta$  & $\sigma =
{(k+\ell-2)\pi \over N}+{\theta\over N}$
\\ & \\
\hline
\end{tabular}
\end{center}
\vspace{0.5cm}
\noindent
Essentially, the off-diagonal matrix elements 
of matrix string are nothing but the 
higher Kaluza-Klein momentum modes, and 
the average value of row and column indices 
indicates the position with respect to the 
$\sigma$ world-volume coordinate. 
The $\sigma$ space of periodicity $2\pi$ is 
decomposed into $N$ segments of length $2\pi/N$.  
Because of this, the integral over $\theta$ together 
with the trace operation can cover the 
spatial world volume of the membrane.  
Note that the segments corresponding to the 
off-diagonal elements $(k, \ell)$ with even 
$k+\ell$ and odd $k+\ell$, respectively,  are shifted by half 
unit $\pi/N$ from each other. 

In the above table, we assumed that the indices $k$ and $\ell$ take 
generic values such that $|k-\ell| \ll N$. 
When $N$ is finite, we have to specify the 
appropriate boundary condition with respect to 
the period of $\theta$ including the boundary region 
of the indices. Only natural choice,  
which is of the form of gauge transformation and is 
consistent with the 
above condition (\ref{bcond}), is
\EQ
A(\theta +2\pi) =S A(\theta)S^{\dagger}, \quad 
S_{k\ell}=\delta^{(N)}_{k+1 \, \, \ell}\, , \, \, (SS^{\dagger}=1, 
\quad S^N=1)
\label{fnboundarycond}
\EN
where  the Kronecker symbol $\delta^{(N)}_{k \, \ell}$ 
should be understood 
as being valid modulo $N$ with respect to 
the matrix indices $k, \ell$.\footnote{
Actually, there is a phase degree of freedom of 
the form $\delta^{(N)}_{k \ell}\exp[(i(\beta_k-\beta_{\ell})
\theta)]$. The phase can however is absorbed by redefining 
the field $A(\theta)$. }
 For generic off-diagonal 
matrix elements, this coincides 
with (\ref{bcond}) and also with the original orbifold 
condition for diagonal elements. 
To preserve the modulo $N$ property exactly 
including the off-diagonal elements, however, 
it becomes necessary to modify the 
assignment of $\sigma$-coordinates such that 
the off-diagonal elements can also be regarded as 
fields on the base $\sigma$-space. For example, 
the above boundary condition (\ref{fnboundarycond}) 
indicates that the matrix element $A_{1N}(\theta)$ with 
shift $\theta\rightarrow \theta +2\pi$ 
is continued to $A_{2\, \, N+1}(\theta) =
A_{21}(\theta)$, the first Kaluza-Klein 
mode $\{A_{k \, \, k+1}(\theta)\} $ in terms of the membrane picture. Similarly, $A_{2N}(\theta)$ 
is continued to $A_{3\,  \, N+1}=A_{3 1}$ which is 
the second Kaluza-Klein mode. 
In general, this shows that the Kaluza-Klein modes are 
cut off such that the KK mode number does not 
exceed $N/2$. 

This modulo $N$ property ($S^N=1$) requires 
the following modifications of the naive 
correspondence explained in the table. 
First,  we require that all the fields be periodic under the 
shift $\sigma \rightarrow \sigma +\pi$, since 
the shift of the matrix indices $k\rightarrow k + N$ or 
$\ell\rightarrow \ell +N$ is 
equivalent to the shift $\sigma \rightarrow \sigma + \pi$. 
In the usual treatments of matrix string theory, 
off-diagonal matrix elements are completely 
neglected except at the interaction points, and 
then the orbifold boundary condition 
would not be important for off-diagonal elements. 
However, when the off-diagonal elements are kept on equal 
footing with the diagonal ones as in the present 
work, the half periodicity is a natural requirement 
in order to satisfy
 the orbifold condition (\ref{fnboundarycond}) 
if we interpret the matrix variables as fields on 
the $\sigma$ space. 
Note that the half period can equivalently 
be formulated as the truncation of the 
Fourier modes with respect to $\sigma$ to 
only even modes.  This is a consistent reduction 
in quantum theory, since it is closed under arbitrary 
algebraic manipulation. Also the quantum states 
of light-cone strings represented by the 
diagonal elements are not reduced at all, 
provided that one 
makes appropriate redefinition of operators both for 
zero and nonzero modes, 
because of global scale invariance remaining in 
light-cone string theory.  Similarly, 
the change of periodicity can also be 
trivially done on the side of membrane, without changing 
the string-theory parameters,  by using the global scale 
invariance of the action with respect to 
two-dimensional base space $(\tau, \sigma)$.

The second modification is that the $\rho$ space 
must be discretized so that we can restrict the 
KK momentum $|n|$ up to $|n| \le N/2$. 
Thus, instead of a circle, the $\rho$ space must be 
assumed to be a Z$_N$ `clock' space, 
$\rho \in [2\pi n/N]\, \,  (n=n+N)$.  In the large 
$N$ limit, the clock space would smoothly 
be approximated by a continuous circle. 
To summarize these two modifications, 
the right hand column in the first line of the above table 
should be understood as 
\EQ
\int_0^{2\pi} d\sigma {1\over 2\pi}\int_0^{2\pi} d\rho
\rightarrow 2\int_0^{\pi} d\sigma {1\over N}\sum_{\rho\in Z_N} .
\EN
Correspondingly, the right hand column in the third line should also be modified to 
\EQ
\int {d\rho\over 2\pi}  e^{-in\rho} A(\sigma, \rho)
\rightarrow {1\over N}\sum_{\rho\in Z_N}e^{-in\rho} A(\sigma, \rho)  .
\EN

With these modifications, the 
appropriate assignment of the $\sigma$-coordinate, 
corresponding to the treatment of the 
off-diagonal elements as Kaluza-Klein fields 
on the base $\sigma$-space,  can be
 given as follows:

\vspace{0.3cm}
\noindent 
When $N=$odd, 
\EQ
\sigma =\left\{
\begin{array}{ll}
{(k+\ell-2)\pi\over N}+{\theta\over N}, & |k-\ell|\le {N-1\over 2} \\
& \\
{(k+\ell-N-2)\pi \over N} +{\theta\over N}, & |k-\ell|>{N-1\over 2}
\end{array}
\right.
\EN

\noindent When $N=$even, 
\EQ
\sigma =\left\{
\begin{array}{ll}
{(k+\ell-2)\pi\over N}+{\theta\over N}, & |k-\ell|\le {N\over 2} \\
& \\
{(k+\ell-2)\pi\over N}+{\theta\over N}, & k-\ell= {N\over 2} \\
&\\
{(k+\ell-N-2)\pi\over N}+{\theta\over N}, & k-\ell=-{N\over 2}\\
&\\
{(k+\ell-N-2)\pi \over N} +{\theta\over N}, & |k-\ell|>{N\over 2}
\end{array}
\right.
\EN

\vspace{0.3cm}
\noindent
We have used the modulo $N$ property in translating the 
$\sigma$ coordinate by $\pi$ for the off-diagonal matrix elements 
with $|k-\ell|\ge N/2$. 
In the following, in order 
to avoid unnecessary complications,  we always use the 
{\it notation} of the naive correspondence. 
But for finite $N$, all the above modifications must be 
tacitly assumed. 

Under these caveats, we can now establish the 
following general formula between the 
integral of traces of matrix string variables  and 
the corresponding membrane variables. 
\EQA
&&{1\over N}\int d\theta \, \Tr\big(
M^{(1)}(\theta)M^{(2)}(\theta) \cdots 
M^{(\ell)}(\theta)\big)
\label{geneformula1} 
\\
&&\hspace{-0.8cm}
={1\over 2\pi}
\int d\rho\int d\sigma 
\exp \Big[ -i{\pi\over N}\sum_{\ell\ge i>j\ge 1}
(\partial_{\sigma_j}\partial_{\rho_i}
-\partial_{\rho_j}\partial_{\sigma_i}) \Big]
M^{(1)}(\sigma_1, \rho_1)\cdots  
M^{(\ell)}(\sigma_{\ell},\rho_{\ell})
\Big|_{\sigma_i =\sigma, \rho_i=\rho} .
\nonumber
\EQN

In order to prove this, 
let us first consider a product of two arbitrary 
matrix-string fields $A(\theta), B(\theta)$. 
\[
(AB)_{k\ell}(\theta)
=\sum_m\int {d\rho_1\over 2\pi}
\int {d\rho_2 \over 2\pi} 
e^{-i(k-m)\rho_1}
e^{-i(m-\ell)\rho_2} 
\]
\[
\times 
A(\sigma_{k\, \ell}(\theta), \rho_1)
\exp({(m-\ell)\pi\over N}\lpartial_{\sigma})
\exp({(m-k)\pi\over N}\rpartial_{\sigma})
B(\sigma_{k\, \ell}(\theta), \rho_2) .
\]
Here the exponential differential operators appeared 
to take account of the difference of the 
$\sigma$ coordinates of the two fields,  
and the $\sigma_{k \, \ell}(\theta)$ is the 
$\sigma$-coordinate determined by the 
indices $k, \ell$ as above. 
By making a change of integration variables 
$(\rho_1, \rho_2) 
\rightarrow (\rho=\rho_1, \tilde{\rho}=\rho_2-\rho_1)$, 
this can be rewritten as 
\[
(AB)_{k\ell}(\theta)
=\sum_m\int {d\rho\over 2\pi}
\int {d\tilde{\rho} \over 2\pi} 
e^{-i(k-\ell)\rho}
e^{-i(m-\ell)\tilde{\rho}} 
\]
\[
\times 
A(\sigma_{k\, \ell}(\theta), \rho)
\exp({(m-\ell)\pi\over N}\lpartial_{\sigma})
\exp({(m-\ell)\pi\over N}\rpartial_{\sigma})
\]
\[
\times \exp({(\ell-k)\pi\over N}\rpartial_{\sigma})
\exp({\tilde{\rho} \rpartial_{\rho}})
B(\sigma_{k\, \ell}(\theta), \rho)
\]
Performing the $\tilde{\rho}$ integral (sum) 
which leads to Kronecker $\delta$ constraint 
$m-\ell = -i\partial_{\rho}$, we can eliminate the 
summation over $m$ and  
further make the partial integral (sum) 
over $\rho$, 
we arrive at the formula
\EQ
(AB)_{k\ell}(\theta)=
\int {d\rho\over 2\pi}
e^{-i(k-\ell)\rho}\Big(
A(\sigma_{k\, \ell}(\theta), \rho)
\exp[-{i\pi\over N}(\lpartial_{\sigma}
\rpartial_{\rho} -
\rpartial_{\sigma}\lpartial_{\rho}
)]
B(\sigma_{k\, \ell}(\theta), \rho)
\Big)
\EN
where the derivatives on the exponential 
inside the big bracket act only within the 
bracket. Note that this general expression 
is valid for finite $N$ too under the 
replacements explained before. In particular, 
the exponential operator 
$\exp[-{i\pi\over N}(\lpartial_{\sigma}
\rpartial_{\rho} -
\rpartial_{\sigma}\lpartial_{\rho}
)]
$ and those appeared in the above manipulation 
are all well defined in the discrete clock space $Z_N$ since 
the eigenvalues of $i\partial_{\sigma}$  are even 
integers. Thus the naive correspondence, including 
the correspondence between commutator and Poisson 
bracket 
which motivated our discussion,  is naturally 
extended to finite $N$ theory with modulo $N$ property. 
For example, expansion in $1/N$ trivially gives  
the correspondence (\ref{commpoisson}). 

Now, applying the above product formula to a general multiple 
product, we obtain
\[
\sum_{i_2, i,_3, \ldots, i_{\ell}}M^{(1)}_{i_1i_2}
(\theta)M^{(2)}_{i_2i_3}(\theta) \cdots 
M^{(\ell)}_{i_{\ell}i_{\ell+1}}(\theta)=
\]
\EQ
\int {d\rho\over 2\pi}
e^{-i(i_1-i_{\ell+1})\rho}
\exp \Big[- i{\pi\over N}\sum_{\ell\ge i>j\ge 1}
(\partial_{\sigma_j}\partial_{\rho_i}
-\partial_{\rho_j}\partial_{\sigma_i}) \Big]
M^{(1)}(\sigma_1, \rho_1)\cdots  
M^{(\ell)}(\sigma_{\ell},\rho_{\ell})
\Big|_{\sigma_i =\sigma, \rho_i=\rho} ,
\EN
where $\sigma=\sigma_{i_1 + i_{i+1}}$.
Taking  trace of this expression 
implies $i_1=i_{\ell+1}$ and 
$\sum_{i_1=i_{\ell+1}}\int d\theta/N
=\int d\sigma$, so that we arrive 
at the promised formula 
(\ref{geneformula1}). 
Although the formula does not look manifestly cyclically
symmetric, we can easily prove cyclic 
symmetry using partial integrations 
$\sum_{i=1}^{\ell}\partial_{\rho_i}=
\sum_{i=1}^{\ell}\partial_{\sigma_i}=0 $:
Indeed, the exponentiated differential 
operator in the formula can be replaced by 
\EQ
\exp \Big[ -i{\pi\over N}\sum_{\ell\ge i>j\ge 2}
(\partial_{\sigma_j}\partial_{\rho_i}
-\partial_{\rho_j}\partial_{\sigma_i}) \Big]
=
\exp \Big[- i{\pi\over N}\sum_{\ell-1\ge i>j\ge 1}
(\partial_{\sigma_j}\partial_{\rho_i}
-\partial_{\rho_j}\partial_{\sigma_i}) \Big]
\EN
which allows us to cyclically change the positions 
of the matrices located at ends inside the trace. 

We note that the above formulas are very similar to the 
well known Moyal product but are not identical: 
The similarity comes from the resemblance of our identification of KK modes and $\sigma$-coordinate 
with matrix indices to 
the Wigner  representation of matrices. 
The difference comes from 
our asymmetrical treatment of the base space coordinates  $\sigma$ and $\rho$, 
in that the former is combined with the 
base-space coordinate $\theta$ of matrix fields 
while the latter is not 
(and is discretized for finite $N$).  
Note that the combination of $\sigma$ and $\theta$ 
is directly responsible to the 
orbifold condition on the matrix-string side. 
 
Now, using the above formula,  we can derive, for instance,  the correspondence 
\[
{1\over N}\int d\theta \, {\rm STr}
\Big([M^{(1)}(\theta), M^{(2)}(\theta)]
[M^{(3)}(\theta), M^{(4)}(\theta)]M^{(3)}(\theta)
\cdots M^{(\ell)}(\theta)\Big)
\]
\[
=
-{1\over 2\pi}(2\pi/N)^2 
\int d\sigma d\rho\, 
\{M^{(1)}(\sigma,\rho), M^{(2)}(\sigma, \rho)\}
\{M^{(3)}(\sigma, \rho), M^{(4)}(\sigma, \rho)\}
\]
\EQ
\hspace{0.8cm}\times  M^{(3)}(\sigma, \rho)
\cdots M^{(\ell)}(\sigma, \rho)
(1+O(1/N^2))
\label{geneformula2}
\EN 
in the large $N$ limit, 
where the symmetrized trace (STr) means to 
treat the commutators in the left hand side 
as single matrices.  The fact that the correction is 
of order $O(1/N^2)$ is owing to  the 
symmetrized trace. 
Using this and similar formulas, we can convert 
arbitrary terms of the membrane action into 
the corresponding ones of matrix string theory in the 
large $N$ limit. 
Thus it is clear that most symmetry 
properties of the supermembrane theory 
are also valid in the matrix-string representation 
interpreted in our way up to the order $O(1/N^2)$   
corrections, provided that the 
symmetry transformation can consistently 
be expressed using the matrix-string degrees of freedom. 
Furthermore, when the 
theory is extended to various nontrivial backgrounds, 
the corresponding symmetries should also be 
ensured in similar ways.

It is now straightforward to map the 
supermembrane action into matrix representation 
by using the established correspondence. 
Using (\ref{geneformula1}) (in particular 
(\ref{geneformula2})), the membrane action  (\ref{doubledimmemaction}) 
is rewritten as, up to $O(1/N^2)$ corrections, 
\EQA
A={(2\pi)^2L\over \ell_M^3}&&\hspace{-0.7cm}
\int d\tau {2\pi \over N}\int_0^{2\pi}
d\theta \, 
\Tr\Big(
{1\over 2}F_{0, \theta}^2+
{1\over 2}(D_0X^i)^2 -{1\over 2}N^2(D_{\theta}X^i)^2
+{1\over 4L^2}({N\over 2\pi})^2[X^i, X^j]^2
  \nonumber \\
&&+i\psi^T D_0\psi -Ni\, \psi^T\Gamma_9D_{\theta}\psi
-{1\over L}{N\over 2\pi}\psi^T\Gamma_i[X^i,\psi]
\Big) ,
\label{mataction}
\EQN
where 
\EQ
D_{\theta}X^i =\partial_{\theta}X^i-i{1\over 2\pi L}[Y, X^i] ,
\EN
\EQ
D_0 X^i =\partial_{\tau}X^i-i{N\over 2\pi L}[A, X^i] ,
\EN
\EQ
F_{0,\theta}
=\partial_{\tau}Y -N\partial_{\theta}A
-i{N\over 2\pi L}[A, Y] ,
\EN 
and similarly for fermion variables. 
By performing the redefinition 
\[
\tau \rightarrow \tau/N, \quad
L\rightarrow L/2\pi, \quad \psi \rightarrow \sqrt{N}\psi,  
\] 
the $N$ dependence is eliminated and the action is 
reduced to  the standard matrix-string theory action. 
Assuming that the physical light-cone time 
$X^+$ must be independent on $N$, 
this rescaling of the time coordinate requires, 
by the relation (\ref{timerelation}),  that 
the total longitudinal momentum $P^+$ scales 
with $N$ as 
\EQ
P^+ \rightarrow NP^+
\label{pplus}
\EN
which coincides with the correct scaling for the 
matrix-string theory interpretation. 
Namely, the diagonal matrix elements of 
$X^i$ represent a fundamental string bit in the 
large $N$ limit and 
consequently the total longitudinal momentum 
is proportional to the number $N$ of the string bits.\footnote{
As discussed in the Appendix 
of the first reference in \cite{dvv-motl}, 
the normalization of the matrix-string 
action in the convention where explicit 
$N$ dependence is completely eliminated gives 
$P^+ \propto N$  in agreement with (\ref{pplus}). }

In this way, we have succeeded to derive the 
matrix-string theory in the large $N$ limit 
directly  from the 
supermembrane action. 
One might wonder what is the relation 
of this method with that based on the standard method 
\cite{taylor} of compactifying general  matrix models of 
D-branes. The above result shows that 
each segment, parametrized by 
$\theta$, of the $\sigma$-space  corresponds 
to the collection of an infinite number of 
image space of a D0-brane. 
The $\theta$ parameter is nothing but 
the conjugate coordinate to the winding number of 
the image spaces. At this point, we 
would like to remind the reader of the fact that 
usual matrix models of D-branes should however 
be regarded as the effective low-energy 
description of D-branes keeping only 
the lowest modes of open strings. 
Then, results obtained by the standard prescription 
should also be regarded as being rigorously 
valid only in some special situation such as in the  
DLCQ limit, where 
the low-energy approximation can be trusted.  
In contrast to this, our method is basically 
independent of any such assumptions, and 
hence it seems reasonable to expect 
more general applicability of our method 
than the standard approach at least in the 
context of establishing connection 
between matrix string and membrane. 

We note also that 
our identification of the off-diagonal 
components of matrix strings 
with the Kaluza-Klein modes of membrane 
along the 
compacfied direction is consistent with the 
usual approach based on T- and S-dualities, 
which suggests that the off-diagonal components 
would correspond to the fluctuating fields 
associated with bound states 
of D0 and F1 string bits.\footnote{
To our knowledge, there has been no literature discussing  explicitly 
the physical meaning of the off-diagonal 
components of matrix string theory. 
Only reference which is related to this question seems to be 
\cite{ghv}, where it is shown that 
the one-loop quantum 
fluctuation 
of the off-diagonal components 
leads to D0-$\overline{\mbox{D0}}$ creation. } Indeed, D0 charge 
is carried by the Kaluza-Klein momentum along 
the compactified M-theory. If we assume further that 
each string bit can only carry the smallest unit of 
Kaluza-Klein momentum, the  mass of 
the fluctuating field should be 
proportional to $|\partial_{\sigma}x|/g_s$, which is 
indeed the case as exhibited in  
(\ref{diaoffdiabracket}).

In the language of matrix string theory, the 
interaction of strings in the limit of small 
$g_s$ has been shown to be understood as 
resulting from the world-sheet instanton effect \cite{ghv} 
where the coincident diagonal matrix elements 
are permuted. By our mapping between the membrane 
picture and matrix-string picture, the 
singular topology change of the strings 
corresponding to the vanishing of $\partial_{\sigma}x^i$ 
in the membrane picture is now 
mapped into this instanton effect 
in the matrix-string picture. 
However, we have to keep in mind that the same 
difficulty as we have discussed 
in connection with quantum-mechanical 
double-dimensional reduction of membrane 
still remains in reducing the 
matrix string action to the light-cone 
string action by integrating over the 
off-diagonal matrix variables. 
Also, the fact that 
membrane should  actually be interpreted as 
 the second quantized 
theory  by the matrix representation is 
equally true as in the ordinary matrix regularization. 
As in the latter case, taking the large $N$ limit of 
the matrix string theory should really 
amount to providing a proper way 
of defining the supermembrane theory.  

Finally, we have to recall that, for the existence of 
longitudinal coordinate $X^-$, 
the condition (\ref{cyclecondition}) must be 
imposed along the $\rho$-direction. This will be 
important for discussing the dynamics. 
Keep in mind however  
that, in a finite $N$ approximation, there is no obvious counterpart 
to this condition, since at least apparently 
the Gauss-law constraint 
for finite $N$ cannot be interpreted as the 
integrability condition.

\vspace{0.4cm}
\section{Matrix-string  theory action in 
nontrivial background: An example}
\setcounter{equation}{0}

We have emphasized that one of the possible 
merits of establishing direct correspondence 
between membrane and matrix string is that 
it enables us to write down the matrix-string theory 
in nontrivial backgrounds, since on the 
side of membrane we can in principle know the 
form of the action on arbitrary background which 
satisfies the field equation of 11 dimensional supergravity \cite{bst}. 
Indeed, one of the natural methods of studying 
string theory in the presence of nontrivial RR backgrounds  
has been to start from supermembrane in 11 dimensions 
and to perform classical double-dimensional reduction. 
In this section, as a simplest nontrivial
 application of our method 
to this direction, 
we construct the matrix-string 
action in a RR background which is called 
Kaluza-Klein Melvin 
(flux 7-brane) background. Recently, the
backgrounds of this type have become a focus of some interests in connection to 
the possible duality relation between type 0 and II theories. 
See, {\it e.g.},  \cite{rt1,rt2,cg}. 

The Kaluza-Klein Melvin background in 10 dimensions is obtained
from the flat space-time in 11 dimensions by the compactification
with non-trivial topology.
We pick up  two, say, 7th and 8th,  of the transverse coordinates
 and make the following identification 
mixing them with the compactified 9th coordinate:
\EQ
(r,y,\varphi)\simeq (r,y+2\pi L m, 
\varphi+2\pi qL m+2\pi n)
\label{periodicity}
\EN
where $y\equiv x^9$ and $x^7+i x^8=r\e^{i\varphi}$. 
That is, we combine the $2\pi L$ shift in the $y$-direction with 
the $2\pi qL$ rotation in the $x^7$-$x^8$ plane. 
For our later purpose, it is more convenient to 
define the coordinates
which are single-valued  in the $y$-direction:
\EQ
x_{flat}^7+i x_{flat}^8=\e^{iq y}(x^7+i x^8),\;
\psi_{flat}=\e^{-{q\over 2}\Gamma_{78}y}\psi
\label{coordtrsf}
\EN
where the subscript `flat' denotes the original coordinates of flat
11-dimensions, which after imposition of the periodicity 
condition (\ref{periodicity}) are no more single-valued when 
$qL\ne$ integers (in the case of fermion when 
$qL\ne$ even integers). Because of this, the system 
after this transformation can describe a nontrivial 
curved background. 
The 11-dimensional metric  after this transformation is given as
\EQ
ds_{11}^2=-dt^2+dx_1^2+\cdots +dx_6^2 +dr^2+r^2(d\varphi+qdy)^2
+dy^2+dx_{10}^2.
\EN

Following the standard 
relation between the metric in 11 dimensions
and the string-frame metric, dilaton and RR 1-form in 10 dimensions
\EQ
ds_{11}^2=\e^{-2\phi/3}ds_{10}^2+\e^{4\phi/3}(dx_{11}+A_\mu dx^\mu)^2,
\EN
we observe that the 10-dimensional string-theory 
background corresponding
to  the above 11D metric is given as
\EQA
&&ds_{10}^2=f(r)[-dt^2+dx_1^2+\cdots +dx_6^2 +dr^2+
r^2 f^{-2}(r)d\varphi^2
+dx_{10}^2]\nonumber\\
&&\e^\phi=f^{3/2}(r), \quad A_\varphi=qr^2f^{-2}(r)\nonumber\\
&&f(r)=(1+q^2 r^2)^{1/2}
\EQN
The RR vector field is magnetic, and both 10 D metric 
and the dilaton are nonpolynomial. 
Since the fermion can acquire $-1$ ($\sim$ anti-periodic 
boundary condition along the $y$-direction)  
under the shift of $q$, the meaningful range of the magnetic 
charge $q$ is $-1/L < q \le 1/L$.  Since our purpose here is only to demonstrate 
a simple application of our formalism for nontrivial 
backgrounds, we assume the ordinary periodic 
boundary condition for spinor field $\psi$. 
To treat the case of antiperiodic boundary condition, 
it is necessary to 
modify our prescription appropriately. 
We will touch upon this only very briefly 
in the end of this section.

The matrix-string action 
in this background is obtained from the
supermembrane action in the flat background, simply  
by rewriting it in terms of the new rotated (single-valued) fields 
via (\ref{coordtrsf}) and by
using the correspondence established in the last section.\footnote{
In preparing the present manuscript, we became 
aware of the work \cite{motl2} which 
discusses a matrix-string 
version of the Kaluza-Klein Melvin 
background from the viewpoint of 
the standard DLCQ approach  to the  
compactification of Matrix models.  We hope that 
our treatment provides a complementary 
account to this important problem. }

We start from the flat space action
\EQ
A=\int d\tau d\sigma d\rho \, \, \Big[{1\over 2}(D_0X^a)^2 
+i\psi^TD_0\psi-{1\over 4}\{X^a,X^b\}^2
+i\psi^T\Gamma_a\{X^a,\psi\}\Big]
\EN
where we have already redefined the normalization of 
$\psi$ from
(\ref{lightconemembrane}). Also note that in this section,
we set the length scales $ L=1, 2\pi\alpha'=1$ for simplicity 
of formulas. 
Now by applying the transformation (\ref{coordtrsf}), 
 the action in terms of the rotated 
fields is given   by replacing the world volume 
derivatives $\partial_\alpha$ $(\alpha=\tau,\sigma,\rho)$ 
with the following `covariant derivatives' $\nabla_\alpha$
\EQA
&&\nabla_\alpha X^m=\partial_\alpha X^m
+q\partial_\alpha Y \epsilon^{mn} X^n,
\nonumber\\
&&\nabla_\alpha X^i=\partial_\alpha X^i,\;
\nabla_\alpha Y=\partial_\alpha Y,\nonumber\\
&&\nabla_\alpha \psi=\partial_\alpha \psi -{q\over 2}\partial_\alpha Y
\Gamma_{78}\psi
,
\EQN
where the indices $i$ denote the `trivial' transverse directions
($i=1,\ldots, 6$) and the directions where rotation
is performed are indicated by $m,n,p=7,8$.  The action
reads
\EQA
A&=&\int d\tau d\sigma d\rho 
\, \, 
\Big[{1\over 2}(D_0 Y)^2 +{1\over 2}(D_0 X^{i})^2
+{1\over 2}(\nabla_0 X^{m}-\{A,X^m\}_\nabla)^2 
-{1\over 2}\{Y,X^{i}\}^2\nonumber\\
&&-{1\over 2}\{Y,X^m\}_\nabla^2 
-{1\over 4}\{X^{i},X^{j}\}^2
-{1\over 2}\{X^{i},X^{m}\}_\nabla^2
-{1\over 4}\{X^{m},X^{n}\}_\nabla^2
+i\psi^T \nabla_0 \psi\nonumber\\
&& -i\psi^T \{A, \psi\}_\nabla
+i\psi^T \Gamma_{i}\{ 
X^{i}, \psi \}_\nabla
+i\psi^T \Gamma_{9}\{ 
Y, \psi \}_\nabla
+i\psi^T \Gamma_m \{ 
X^m, \psi \}_\nabla \Big]
,
\EQN
where subscript $\nabla$ means that the Poisson bracket
is evaluated using the above covariant derivatives.
The action has the order $O(q^0)$-, $O(q^1)$- and $O(q^2)$- parts.
The $q^0$-part
is of course 
the same as the original action except for the fact
that the fields are now the redefined ones.
The $q^1$- and $q^2$- parts are given as
\EQA
A^{q^1}&=&q\int d\tau d\sigma d\rho \, \epsilon^{mn}
\Big[-D_0YD_0X^{m} X^{n} + \{X^{i},Y\}\{X^{i},X^{m}\}X^{n}\nonumber\\
&&+\{X^{p},Y\}\{X^{p},X^{m}\}X^{n}
-i\psi^T \Gamma_{m} X^{n}\{Y,\psi\}
\nonumber\\
&&-{i\over 4}\psi^T \Gamma_{mn}\psi D_0 Y
-{i\over 4}\psi^T \Gamma_{i}\Gamma_{mn}\psi 
\{ X^{i}, Y \}\Big]
, 
\label{aq1}
\EQN
\EQ
A^{q^2}=q^2\int d\tau d\sigma d\rho 
\, \Big[{1\over 2}(D_0 Y)^2 (X^{m})^2 -{1\over 2}\{X^{i},Y\}^2
(X^{m})^2 -{1\over 2}\left( X^{m}\{X^{m},Y\}\right)^2
\Big]
\label{aq2}
\EN
Here we have directly applied the coordinate 
transformation to the light-cone action. 
However, it is easy to check that 
we obtain the same result if we first make 
the coordinate transformation and afterwards 
go to the light-cone gauge.  The covariant action resulting from the latter
procedure is in fact given in \cite{rt1}.
We have explicitly checked that the Hamiltonian 
obtained from
our procedure ({\it i.e.} using the action (\ref{aq1}), 
(\ref{aq2})) in the $A_0=0$ gauge agrees with the light-cone-gauge
Hamiltonian obtained from the covariant action of \cite{rt1}. 
This is as expected since the coordinate 
rotation is performed
in the transverse directions, so the light-cone gauge fixing and
the rotation should commute. 
We also note that 
when the gauge field $A_0$ is integrated over, 
the dependence on the background charge 
$q$ becomes nonpolynomial.

To study the compactified membrane, we perform the shift 
$Y\rightarrow \rho +Y$ in the above action.
As we have seen in section 2, after the substitution,
$Y$ plays the the role
of the $\sigma$-component of 2-dimensional gauge field. 
The resulting order $O(q)$ and $O(q^2)$ actions are 
\EQA
A^{q^1}&=&q \int d\tau d\sigma d\rho\,\epsilon^{mn} 
\Big[-F_{0,\sigma}D_0 X^{m} X^{n} 
+D^Y_\sigma X^{i}\{X^{i},X^{m}\}X^{n}\nonumber\\
&&+D^Y_\sigma X^{p}\{X^{p},X^{m}\}X^{n}
+i\psi^T \Gamma_{m} X^{n} D^Y_\sigma\psi
\nonumber\\
&&-{i\over 4}\psi^T \Gamma_{mn}\psi F_{0,\sigma}
-{i\over 4}\psi^T \Gamma_{i}\Gamma_{mn}\psi D^Y_\sigma X^{i}\Big]
 , 
\label{sq1}
\EQN
\EQ
A^{q^2}=q^2 \int d\tau d\sigma d\rho 
\,  \Big[{1\over 2}(F_{0,\sigma})^2 (X^{m})^2 
-{1\over 2}(D^Y_\sigma X^{i})^2 (X^{m})^2
-{1\over 2}(X^{m}D^Y_\sigma X^{m})^2\Big]
\label{sq2}
\EN
where $D^Y_\sigma$ and $F_{0,\sigma}$ are those 
defined in 
section 2 (with $L=1$).

Now we follow the correspondence between compactified 
membrane and 
Matrix string and obtain the Matrix-string 
representation of the
action. The dictionary of the correspondence is given 
in the table in section 3. Especially, the Poisson brackets of 
membrane fields correspond to the commutators of matrices
in the leading order in the large $N$ limit.
Also, the orbifold boundary condition 
(\ref{fnboundarycond}) must be kept in mind. 
$q$-independent part of the action is quadratic in the fields
when we regard Poisson brackets as a single unit, and it was shown
in section 3 that this part is mapped to 
the Matrix-string action
in the flat background.
The order $q^1$-part and $q^2$-part are cubic and quartic in the fields
respectively, treating the Poisson bracket as a single unit.
Using the general formula (\ref{geneformula1}), we can easily see
that to the leading order in the large $N$ limit up to 
$O(1/N^2)$ correction, they are expressed 
using the symmetrized trace as follows.
\EQA
A^{q^1}&=&q \int d\tau d\theta\, \epsilon^{mn}\,{\rm STr}
\Big[-F_{0,\theta}D_0 X^{m} X^{n} +i D_\theta
X^{i}[X^{i},X^{m}]X^{n}\nonumber\\
&&
+i D_\theta X^{p}[X^{p},X^{m}]X^{n}
+i\psi^T \Gamma_{m} X^{n} D_\theta\psi
\nonumber\\
&&-{i\over 4}\psi^T \Gamma_{mn}\psi F_{0,\theta}
-{i\over 4}\psi^T \Gamma_{i}\Gamma_{mn}\psi 
D_\theta X^{i} \Big]
\EQN
\EQ
A^{q^2}=q^2 \int d\tau\int d\theta\, {\rm STr}
\Big[{1\over 2}(F_{0,\theta})^2 (X^{m})^2 
-{1\over 2}(D_\theta X^{i})^2 (X^{m})^2
-{1\over 2}(X^{m}D_\theta X^{m})^2 \Big]
\EN
where we have performed the rescaling as in the 
flat case. The definitions of
covariant derivatives and field strength are as follows.
\[
D_{\theta}X^i =\partial_{\theta}X^i-i[Y, X^i],\quad
D_0 X^i =\partial_{\tau}X^i-i[A, X^i] ,
\]
\[
F_{0,\theta}
=\partial_{\tau}Y -\partial_{\theta}A
-i[A, Y].
\] 

In the linearized approximation where 
we ignore  the $O(q^2)$ terms, we can derive 
equivalent results  as ours using the membrane vertex operator 
or matrix-string vertex operator as studied 
in refs. \cite{dnp} or \cite{scia}, respectively, 
{\it provided}  we perform a field redefinition 
of the space-time spinor field corresponding to a
change of local Lorentz frame.  In contrast to the 
linearized approximation, 
however, our result should be valid to all orders in
$q$. For the sake of future reference and 
also as a consistency check, 
we briefly describe the correspondence of our result with the linearized 
approximation in Appendix B.

Finally, let us briefly touch upon the case where the 
transformed spinor $\psi$ is antiperiodic along the 
$\rho$ direction. In this case, 
we need to introduce half-integer Kaluza-Klein 
modes for spinors. The Fourier decomposition 
\EQ
\psi(\sigma, \rho) =\sum_n \psi_{n+1/2}(\sigma)\e^{i(n+1/2)\rho}
\EN
and the Poisson bracket between the zero mode 
coordinate
\EQ
\{x(\sigma), \psi(\sigma, \rho)\}=i
\sum\partial_{\sigma}x(\sigma)(n+{1\over 2})\psi_{n+1/2}(\sigma)
\e^{i(n+1/2)\rho}
\label{antiperibra}
\EN
suggest that a natural extension of our procedure 
discussed in section 3 is to double the range of matrix indices 
$N\rightarrow 2N$ and to assume that the bosonic 
matrix variables have only even-even and odd-odd 
elements while the spinor matrix variables 
have odd-even (or even-odd) elements. Thus the 
number of (real) matrix components for bosons is 
2$N^2$. As  $2N\times 2N$ matrices, 
the off-diagonal matrix elements of 
bosonic matrix-string fields are nonzero only for 
`even' off diagonal lines with the differences between 
row and column being restricted to even integers. 
Correspondingly, the fermion matrices 
are now assumed to be complex $N\times N$. Let us denote 
even integers by $m,n, \ldots$ and odd ones by 
$p, q, \ldots$. 
We associate the integer KK mode numbers with 
differences among $(m, n, \ldots)$ and among 
$(p, q, \ldots)$, while the half-integer KK modes numbers  
with differences between $(m, n, \ldots)$ and $
(p, q, \ldots)$.  
Then the correspondence of matrix-string field and 
membrane field for spinors is 
\EQ
\psi_{p, m}(\theta) \leftrightarrow \psi_{(p-m)/2}(\sigma)
\EN
with the hermiticity condition $\psi^{\dagger}_{p,m}
=\psi_{m,p}$.   As before the
correspondence  of the $\sigma$ and $\theta$ coordinates is 
$\sigma =(p+m-2)\pi/2N + \theta/2N$ ($p, m$ modulo $2N$). 
The bracket relation (\ref{antiperibra}) can then 
be naturally interpreted as a commutator between 
bosonic and spinor matrices, 
for the derivative $(n+1/2)\partial_{\sigma}x$  
can be interpreted as the difference between 
the odd-odd and even-even diagonal matrices of bosons. 

The gauge symmetry is now U($N$)$\times$U($N$) 
under which the 
bosonic matrix variables transform
as adjoint representation, 
while the spinors transform
 as the bifundamental $(\overline{N}, N)$ 
representation. Note also that the 
boundary condition with respect to $\theta 
\rightarrow \theta +2\pi$ connects  
the even-even and odd-odd matrices in the 
bosonic case, while in the fermionic case 
it connects even-odd to odd-even. 
The appearance of similar gauge 
structure for type 0A or Melvin background has 
previously been pointed out in
\cite{banksmotl, motl2} within the standard approach to matrix string. 
In our approach, the theory is completely local 
on the world volume because of our transformation (\ref{coordtrsf}) 
to the single-valued (double-valuded for antiperiodic fermions)  world-volume fields at the level 
of the membrane theory.  
Although we do not elaborate along this direction further 
in the present paper, it would be an interesting problem to 
investigate membrane and matrix strings in the Kaluza-Klein Melvin background using our approach from the viewpoint 
of duality of type 0 and II theories.

\section{Concluding remarks}

In the present work, we have developed a 
method of directly 
mapping the theory of supermembrane wrapped 
along a circle to matrix string theory 
and discussed its implications and applications. 
We hope that our observations may provide 
new impetus toward further exploration of 
the dynamics of membranes and matrix strings.  
Here we mention some remaining problems 
and future possibilities.  

An interesting possibility is to 
apply our method to the 
covariant treatments of (super) membrane. 
For a relatively recent attempt of covariant quantization 
of membrane in 10 dimensional sense, 
see {\it e.g.} ref. \cite{okufuji}. Unfortunately, 
the method of the latter reference does not seem convenient 
for double compactification, because of their 
gauge choice for fermion degrees of freedom. 
If this were successfully done, we would have 
a covariant version of matrix string theory. 
It is also desirable to prove Lorentz invariance 
directly using the matrix-string language. 

In connection to covariantization, another 
intriguing question might be whether our procedure 
of making correspondence between Poisson bracket and matrix  
commutator can be extended usefully 
to higher bracket structures  
such as the Nambu bracket, which seems to 
be relevant 
\cite{awaliminiyone} 
for fully covariantized formulation of membranes. 

Besides these and other possibilities of extending our 
formalism, one of the most crucial problems related to 
our work  is to study the possibility of generating 11 dimensional 
gravity {\it dynamically} from supermembrane or matrix string. 
Since our correspondence provides a clear 
physical picture for matrix string variables including the 
off-diagonal elements 
from the viewpoint of 11 dimensional membrane theory, 
we may proceed to explore the behavior of D-particles 
using membrane-matrix picture in the 
decompactification limit $L\rightarrow \infty$. 
For this direction, it seems
important  to formulate the detailed dynamical properties of D-particles 
in our membrane-matrix-string approach. 

We hope to return to some of these issues in the 
near future.

\vspace{0.4cm}
\noindent
Acknowledgements

Part of this work was done during participation 
of one (T. Y.) of the present authors in the M-theory workshop at ITP, University of California, Santa Barbara. 
He would like to thank the organizers of the workshop 
for enabling his visit and for 
providing stimulating atmosphere. 
The present work is supported in part by Grant-in-Aid for Scientific 
Research (No. 12440060)  from the Ministry of  Education,
Science and Culture. 

\vspace{0.6cm} 
\noindent
{\large Appendix A}
\renewcommand{\theequation}{A.\arabic{equation}}
\appendix 
\setcounter{equation}{0}
\vspace{0.3cm}
\noindent

 \noindent

In this Appendix, we report preliminary results of a 
quantum mechanical study
on the double dimensional reduction.
It is believed that  the
supermembrane in 11 dimensions wrapped around a circle
becomes the type IIA superstring in 10 dimensions by a simultaneous
dimensional reduction of the world-volume and space-time.
In \cite{dhis}, this was shown classically, {\it i.e.} by
simply discarding the dependence of the fields 
on the compact direction. 
Let us  try to give a justification to this picture by
studying the quantum effective action.  
Contrary to the classical argument, 
we keep the Kaluza-Klein modes on the membrane world-volume
and integrate them out. We analyze the behavior in the
limit of small compactification radius by a strong coupling 
expansion and see whether 
quantum corrections affect the 
string action.

Start from light-cone supermembrane action 
\EQA
A={(2\pi)^2/\ell_M^3}&&\hspace{-0.7cm}
\int d\tau \int_0^{2\pi}
d\sigma
\int_0^{2\pi L} d\rho \Big(
{1\over 2}(D_0X^i)^2+{1\over 2}(D_0Y)^2 
-{1\over 4}\{X^i, X^j\}^2
-{1\over 2}\{X^i, Y\}^2 \nonumber \\
&&+i\psi^T D_0\psi
+i\psi^T\Gamma_i\{X^i,\psi\}
+i\psi^T\Gamma_9\{Y, \psi\}
\Big) .
\EQN
To study the compactified membrane, we set the background of $Y$
equal to the world volume spatial direction $\rho$.
\EQ
Y(\sigma,\rho,\tau)\rightarrow \rho +Y(\sigma, \rho, \tau) , 
\label{ydecomp}
\EN
\EQ
Y(\sigma, \rho, \tau)=
\sum_{n\ne 0} \e^{in\rho/L}Y_n(\sigma,\tau)
.
\label{fourier1}
\EN
Other fields are decomposed into $\rho$-independent and 
$\rho$-dependent parts.
\EQA
&&X^i(\sigma, \rho,\tau)\rightarrow
x^i(\sigma, \tau) +X^i(\sigma,\rho,\tau),\quad 
\psi(\sigma,\rho,\tau) \rightarrow
\psi(\sigma, \tau) +\Psi(\sigma, \rho,\tau) ,
\nonumber\\
&&A(\sigma,\rho,\tau) \rightarrow  a(\sigma, \tau) +
A(\sigma,\rho, \tau) 
,
\label{adecomp}
\EQN
where $X$, $A$ and $\Psi$ are periodic in $\rho$:
\EQA
&&X^i(\sigma, \rho, \tau)=
\sum_{n\ne 0} \e^{in\rho/L}X^i_n(\sigma,\tau),\quad
\Psi(\sigma, \rho,\tau)=
\sum_{n \ne 0}\e^{in\rho/L}\Psi_n(\sigma, \tau)
, \label{fourier2}\\
&&A(\sigma, \rho,\tau)=\sum_{n\ne 0}
\e^{in\rho/L}A_n(\sigma,\tau)\nonumber. 
\EQN
We can treat the $\rho$-independent part as background and
$\rho$-dependent part as fluctuations, 
since the $\rho$-dependent part starts  
from quadratic order. 
Classical action for the backgrounds is nothing but the type IIA string 
action in 10 dimensions.

The most convenient gauge choice  for the fluctuations
($A$, $X^i$, $Y$) is the standard background field gauge condition:
\EQ
\partial_0 A-\{a,A\}+\{x^i, X^i\}+\{\rho ,Y\}=0.
\label{bgfgauge}
\EN
This condition remains 
invariant if the gauge parameter is 
independent of $\rho$, for which the background and fluctuations
transform as
\EQA
&&\delta a=\partial_0 \lambda,\; \delta x^i=0,\; 
\delta \rho=\partial_\sigma \lambda,\\
&&\delta A=\{\lambda, A\},\; \delta X^i=\{\lambda, X^i\},\;
\delta Y=\{\lambda, Y\}. 
\label{fluctrsf}
\EQN
Note that
this residual gauge freedom is used to bring
$Y$ to the form (\ref{ydecomp}).
The gauge-fixing term and the ghost action for the background field gauge
are  as follows.
\EQ
{\cal L}_{{\rm gf}}
=-{1\over 2}(\partial_0 A -\{a,A\} +\{x^i, X^i\}  
-\partial_{\sigma}Y)^2 ,
\EN
\EQA
{\cal L}_{{\rm gh}}&=&
-\partial_0\overline{C}(\partial_0 C -\{a, C\}-\{A, C\})
+\{a,\overline{C}\}(\partial_0 C -\{a, C\}-\{A, C\})
\nonumber \\ 
&&
+\partial_{\sigma}\overline{C}(\partial_{\sigma}C -\{Y, C\})
+\{x^i, \overline{C}\}(\{x^i, C\}+\{X^i, C\}) ,
\EQN
where $C$ and $\overline{C}$ are also periodic in $\rho$
\EQ
C(\sigma, \rho, \tau)=
\sum_{n\ne 0} \e^{in\rho/L}C_n(\sigma,\tau),\quad
\overline{C}(\sigma, \rho,\tau)=
\sum_{n \ne 0}\e^{in\rho/L}\overline{C}(\sigma, \tau) .
\label{fourier3}
\EN

Now to study the small compactification limit $L\rightarrow 0$,
we make a rescaling
\EQ
\rho\rightarrow L\rho.
\EN
As mentioned in section 2, $1/L$ plays the
role of the gauge coupling.
We compute the effective action in expansion of $L$,
which is essentially the strong coupling expansion.
We use the Euclidean formulation
\EQ
\tau\rightarrow -i\tau, a\rightarrow i a, A\rightarrow i A.
\EN
The fluctuations are also rescaled  as
\EQ
A\rightarrow LA, \, 
X^i \rightarrow LX^i, \, Y\rightarrow LY, 
\, \Psi \rightarrow \sqrt{L}\Psi,
C \rightarrow LC.
\label{rescaling2}
\EN
The action at each order of $L$ is given as follows.
The subscript `$0,1,2$' indicate the order with respect to $L$ and
the  subscript `bg' means the term containing only the background
field. 
First, the parts which contain no spinor
fields ($\psi$, $\Psi$) are  
\EQ
{\cal L}_B={\cal L}_{{B,\rm bg}}+ {\cal L}_{B,0} +L^1{\cal L}_{B,1}
 +L^2{\cal L}_{B,2} ,
\EN
\EQ
{\cal L}_{B,{\rm bg}} = {1\over 2}(\partial_0 x^i)^2
+{1\over 2}(\partial_{\sigma}x^i)^2 +{1\over 2}(\partial_{\sigma}a)^2 ,
 \label{lbbg}
\EN 
\EQA
{\cal L}_{B,0} &=& {1\over 2}\big((\partial_{\sigma} a)^2
+(\partial_{\sigma} x^i)^2\big)(\partial_{\rho}X^j)^2
+{1\over 2}\big((\partial_{\sigma} a)^2
+(\partial_{\sigma} x^i)^2\big)(\partial_{\rho}Y)^2
\nonumber \\ &&
+{1\over 2}\big((\partial_{\sigma} a)^2
+(\partial_{\sigma}x^i)^2\big)
(\partial_{\rho}A)^2 -i\big((\partial_{\sigma} a)^2
+(\partial_\sigma x^i)^2\big)\partial_\rho \bar{C}
\partial_\rho C ,
\label{lb0}
\EQN
\EQA
{\cal L}_{B,1} &=&
-2\partial_0 x^i\{A,X^i\}-2\partial_\sigma a\{Y, A\} 
-2\partial_\sigma x^i\{Y,X^i\} \nonumber\\
&&-\partial_\sigma a \partial_0 A \partial_\rho A 
-\partial_\sigma a \partial_0 Y \partial_\rho Y 
-\partial_\sigma a \partial_0 X^i \partial_\rho X^i\nonumber\\
&&-\partial_\sigma a \partial_\rho Y \{Y,A\}
-\partial_\sigma a \partial_\rho X^i \{X^i,A\}\nonumber\\
&&-\partial_\sigma x^i \partial_\rho A \{A,X^i\}
-\partial_\sigma x^i \partial_\rho Y \{Y,X^i\}
-\partial_\sigma x^i \partial_\rho X^j \{X^j,X^i\}\nonumber\\
&&+i\partial_\sigma a\partial_0 \overline{C}\partial_\rho C
+i\partial_\sigma a\partial_\rho \overline{C}\partial_0 C
\nonumber \\
&& +i\partial_\sigma a \partial_\rho \overline{C} \{C,A\}
+i\partial_\sigma x^i \partial_\rho \overline{C} \{C,X^i\} ,
\label{lb1}
\EQN
\EQA
{\cal L}_{B,2} &=&{1\over 2}(\partial_0 A)^2 +{1\over 2}(\partial_0 X^i)^2
+{1\over 2}(\partial_0 Y)^2 +{1\over 2}(\partial_\sigma A)^2
+{1\over 2}(\partial_\sigma X^i)^2+{1\over 2}(\partial_\sigma Y)^2\nonumber\\
&&+\partial_\sigma A\{A,Y\} +\partial_\sigma X^i\{X^i,Y\}
+\partial_0 X^i\{X^i,A\} +\partial_0 Y\{Y,A\}\nonumber\\
&&+{1\over 2}\{A,X^i\}^2+{1\over 2}\{A,Y\}^2+{1\over 2}\{X^i,Y\}^2
+{1\over 4}\{X^i,X^j\}^2\nonumber\\
&&-i\partial_0\overline{C}\partial_0 C -i\partial_\sigma \overline{C}\partial_\sigma
C +i\partial_0\overline{C}\{A,C\}+i\partial_\sigma\overline{C}\{Y,C\}  .
\label{lb2}
\EQN 
The parts containing spinor fields
are 
\EQ
{\cal L}_F={\cal L}_{{F,\rm bg}}+ {\cal L}_{F,0} +L^{1/2}{\cal L}_{F,1/2}
 +L^1{\cal L}_{F,1} ,
\EN
\EQ
{\cal L}_{F,{\rm bg}} =  
\psi^T\partial_0\psi
+i\psi^T\Gamma_9\partial_{\sigma}\psi ,
\label{lfbg}
\EN 
\EQ
{\cal L}_{F,0} = -\Psi^T \partial_{\sigma}a\partial_\rho\Psi
-i\Psi^T\Gamma_i \partial_{\sigma}x^i\partial_\rho\Psi ,
\label{lf0}
\EN 
\EQ
{\cal L}_{F,1/2} = 2\Psi^T \partial_\rho A \partial_\sigma \psi
+2i\Psi^T \Gamma_9 \partial_\rho Y \partial_\sigma \psi
+2i\Psi^T \Gamma_i \partial_\rho X^i \partial_\sigma \psi ,
\label{lfhalf}
\EN
\EQA
{\cal L}_{F,1} &=& \Psi^T\partial_0\Psi 
+\Psi^T\Gamma_9\partial_\sigma\Psi\nonumber\\
&&-\Psi^T\{A, \Psi\} 
-i\Psi^T\Gamma_9\{Y, \Psi\}
-i\Psi^T\Gamma_i\{X^i, \Psi\} .
\label{lf1}
\EQN

We regard the order $L^0$-parts 
as the free action.
Substituting the Fourier expansions (\ref{fourier1}), (\ref{fourier2})
and (\ref{fourier3})
into the action, $L^0$-part in terms of the Kaluza-Klein modes reads
\EQA
A_0&=&{(2\pi)^3 L\over 2\ell_M^3}\int d\tau d\sigma \Big[
n^2 ((\partial_\sigma a)^2+(\partial_\sigma x^i)^2 ) A_n A_{-n}
+n^2 ((\partial_\sigma a)^2+(\partial_\sigma x^i)^2) X^j_n X^j_{-n}
\nonumber\\
&&+n^2 ((\partial_\sigma a)^2+(\partial_\sigma x^i)^2) Y_n Y_{-n}
-2i n^2 ((\partial_\sigma a)^2+(\partial_\sigma x^i)^2) \overline{C}_n C_{-n}
\nonumber\\
&&-2n \Psi_n (\partial_\sigma a +\Gamma_i\partial_\sigma x^i) \Psi_{-n}
\Big] .
\label{s0}
\EQN
The propagators for $X^i_n$, $\Psi_n$ are thus given as
\EQA
&&\langle X^i_{-n}(\xi)X^j_{n}(\xi')\rangle= 
{\delta^{ij}\over n^2}G(\xi,\xi') ,
\nonumber \\
&& \langle \Psi_{-n}(\xi)\Psi_{n}(\xi')\rangle=
{-i\over 2n}(\partial_\sigma a-i\Gamma_i \partial_\sigma x^i)
G(\xi,\xi') ,
\label{propagators}
\EQN
\[
G(\xi,\xi')={1\over (\partial_\sigma a)^2+(\partial_\sigma x^i)^2} 
\delta^{(2)}(\xi- \xi') ,
\]
where $\xi=\tau,\sigma$ and 
we set $(2\pi)^3L/\ell^3_M=1$ for brevity.
Propagators for other fields are given similarly.
We restrict the background configurations where
the U(1) gauge field which have no dynamics is set to zero ($a=0$). 

In our strong coupling expansion, the free part contain no derivatives
of the fluctuations with respect to the world-sheet coordinates
$\tau, \sigma$.  As emphasized in the text, 
this ultra-local action necessarily 
leads to the propagators 
which are proportional to the delta 
function and lead to the UV
divergences $\delta(0)$ upon loop calculations, thus
for a rigorous treatment, we need  a regularization.
Since, as already alluded to in the discussion in section 2,  it seems difficult to find a suitable regularization (which respects
supersymmetry {\it etc.}), we only give a formal and partial
argument for the vanishing quantum correction at low orders in $L$, by demonstrating 
 that the coefficients of the divergence 
vanish after  appropriately 
arranging these singular terms.

First of all, we see from (\ref{s0})
that the lowest order correction, coming from 
logarithm of the one-loop
determinant actually vanishes due to the matching of the bosonic
and fermionic degrees 
of freedom (8 (=10-2) bosonic d.o.f :
$X^i,Y,A,C,\overline{C}$; 8 (=16/2)
 fermionic d.o.f: $\Psi$) . 
We shall see the vanishing of a few low order terms in similar way.

The contribution to the effective action obviously does not have terms of
half-integer order in $L$, for they are associated with odd number of 
fermionic fluctuations.
The effective action at order $L$ would come from $\langle {\cal
L}_{B,1}\rangle$,  $\langle {\cal L}_{F,1}\rangle$ 
and $\langle {\cal L}_{F,1/2}
{\cal L}_{F,1/2} \rangle$. However, 
these three contributions vanish separately 
for the following reasons. 
First, $\langle {\cal L}_{B,1}\rangle=0$
for there is no way to self-contract ${\cal L}_{B,1}$
as we can see from (\ref{lb1}). Also, $\langle {\cal L}_{F,1}\rangle=0$
as we can see from (\ref{lf1}) by noting that 
$\langle \Psi^T\partial_0\Psi\rangle
$$=$$\langle \Psi^T\Gamma_9\partial_\sigma\Psi\rangle$$=0$ 
due to $\Tr \Gamma_i=\Tr \Gamma_i\Gamma_9=0$. 
Finally, $\langle {\cal L}_{F,1/2}
{\cal L}_{F,1/2} \rangle=0$,  for it is proportional to 
$\sum_{n\neq 0}{1\over n}$ which can be set to zero 
by symmetry.

At order $L^2$, possible contributions which do not contain 
background spinor $\psi$ come from 
 $\langle {\cal L}_{B,1}{\cal L}_{B,1}\rangle$, 
$\langle {\cal L}_{B,2}\rangle$
and $\langle {\cal L}_{F,1}{\cal L}_{F,1}\rangle$. 
Each one has one-loop and two-loop terms as shown in the Figure. 
There are ambiguities in the evaluation of the effective action
at this order for we are dealing with the divergent quantities
by formally treating the divergence as $\delta(0)$.
(One-loop terms are proportional to $\delta (0)$ and two-loop
terms are proportional to $(\delta (0))^2$, and hence they 
have different dimensions.  )

\begin{center}
\begin{figure}
\begin{picture}(180,120)
\put(140,0){\epsfxsize 180pt 
\epsfbox{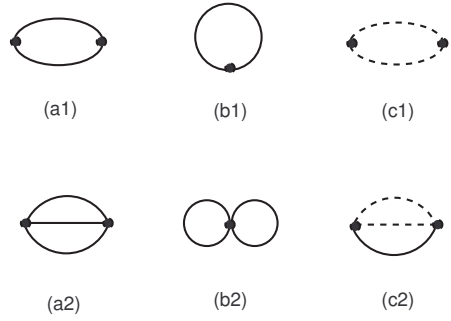}}
\end{picture}
\caption{The order $L^2$ contributions to the effective action. 
(a1),(a2): $\langle {\cal L}_{B,1}{\cal L}_{B,1}\rangle$, 
(b1),(b2): $\langle {\cal L}_{B,2}\rangle$,
and (c1),(c2): $\langle {\cal L}_{F,1}{\cal L}_{F,1}\rangle$, 
respectively. Solid line denotes the propagators 
of bosonic fields or ghosts
and dotted line denotes the propagators 
of spinor fields.} 
\label{Fig1} 
\end{figure}
\end{center}

\vspace{-1.4cm}
We shall show
that the one-loop contributions cancel. 
Explicit form of the one-loop contributions are as follows.
First, from the diagram of Figure (a1),
\EQA
&&\langle {\cal L}_{B,1}{\cal L}_{B,1}\rangle^{(one-loop)}=
-4\int d^2\xi\int d^2 \xi' \sum_{n\neq 0}
{1\over n^2}\nonumber\\
&&\big[
\partial_0 x^i\partial'_0\partial'_\sigma x^i
\partial_\sigma G(\xi,\xi') G(\xi,\xi')
+\partial_\sigma x^i\partial'_\sigma\partial'_\sigma x^i
\partial_\sigma G(\xi,\xi') G(\xi,\xi')\big] ,
\label{a1}
\EQN
where $\partial_\sigma x^i$, $\partial'_\sigma x^i$ means
the argument of $x^i$ is $\xi$, $\xi'$ respectively and similarly
for $\partial_0 x^i$, $\partial'_0 x^i$.
From Figure (b1),
\EQ
\langle {\cal L}_{B,2}\rangle^{(one-loop)}= 
-4 \int d^2\xi\sum_{n\neq 0}{1\over n^2}
\lim_{\xi\rightarrow \xi'} \big[
\partial_0\partial'_0 G(\xi,\xi')
+ \partial_\sigma\partial'_\sigma G(\xi,\xi')\big] .
\EN
We rewrite this term by inserting the delta function
\EQ
\lim_{\xi\rightarrow \xi'}1=\int d^2\xi' \delta(\xi-\xi')
=  \int d^2\xi' \partial_\sigma x^i\partial'_\sigma x^i
G(\xi,\xi')
\label{deltaprop}
\EN
as 
\EQ
\langle {\cal L}_{B,2}\rangle^{(one-loop)}= -4 \int d^2\xi\int d^2\xi'
\sum_{n\neq 0}{1\over n^2}\partial_\sigma x^i\partial'_\sigma x^i
 \big[
\partial_0\partial'_0 G(\xi,\xi')G(\xi,\xi')
+ \partial_\sigma\partial'_\sigma G(\xi,\xi')G(\xi,\xi')
\big] .
\label{b1}
\EN
From figure (c1), 
\EQA
\langle {\cal L}_{F,1}{\cal L}_{F,1}\rangle^{(one-loop)}&=
& 4\int d^2 \xi \int d^2\xi'
\sum_{n\neq 0}{1\over n^2}
\big[
\partial_\sigma x^i\partial'_\sigma x^i G(\xi,\xi')
\partial_0\partial'_0 G(\xi,\xi')\nonumber\\
&&
+\partial_\sigma x^i\partial'_\sigma x^i G(\xi,\xi')
\partial_\sigma\partial'_\sigma G(\xi,\xi')
+\partial_\sigma x^i \partial'_0 \partial'_\sigma x^i G(\xi,\xi')
\partial_0 G(\xi,\xi')\nonumber\\
&&+\partial_\sigma x^i \partial'_\sigma\partial'_\sigma x^i G(\xi,\xi')
\partial_\sigma G(\xi,\xi')
\big] .
\label{c1}
\EQN 

Here, we have to note an ambiguity due 
to a formal treatment of 
divergent quantities.
In writing (\ref{c1}), we have assumed that 
the propagators for spinors are 
related to those of bosons by $\langle \Psi_{-n}(\xi)
\Psi_{n}(\xi')\rangle={-1\over 2n}
\Gamma_i \partial_\sigma x^i(\xi)
G(\xi,\xi')$, but if we have assigned the argument of $x^i$ in 
a different way, ({\it e.g.} $(\partial_\sigma x^i(\xi)
+\partial'_\sigma x^i(\xi'))/2$),
we would have a different answer. 
The sum of the one-loop contributions to the effective action
(\ref{a1}), (\ref{b1}) and (\ref{c1}) vanish
\EQ
\langle {\cal L}_{B,1}{\cal L}_{B,1}\rangle^{(one-loop)}
+\langle {\cal L}_{B,2}\rangle^{(one-loop)}
+\langle {\cal L}_{F,1}{\cal L}_{F,1}\rangle^{(one-loop)}=0.
\EN
To prove this, we have used 
$\partial_\sigma G(\xi,\xi') G(\xi,\xi')=
{1\over 2} \partial_\sigma (G(\xi,\xi') G(\xi,\xi'))$ and
performed partial integration.

We have also checked in a similar way 
that the order $L^2$-contributions
 containing background spinor $\psi$'s 
(two $\psi$'s: $\langle {\cal L}_{F,1/2}
{\cal L}_{F,1/2}{\cal L}_{B,1}\rangle$;
four $\psi$'s: $\langle {\cal L}_{F,1/2}
{\cal L}_{F,1/2}{\cal L}_{F,1/2}{\cal L}_{F,1/2}\rangle$)
vanish essentially due to the antisymmetry of the $\psi$'s.

Two-loop contributions are more ambiguous
and moreover we need an appropriate
 regularization scheme for 
the infinite sum over the Kaluza-Klein level $n$.
This is a difficult question for which 
we do not have a definite answer.  

Almost the same computations can 
be performed for the matrix-string case as well. 
Therefore it is very difficult to 
really justify the 
reduction to 10
dimensions in the infrared limit too. 
However,  vanishing of all these corrections 
seems essential for justifying the 
reduction to diagonal elements and for establishing 
the equivalence of matrix string theory with the perturbative  
superstring theory, such that 
the only remaining effect of the string coupling is 
the usual string interaction which is described by the 
instanton effect corresponding to the 
exchange of coincident eigenvalues of the 
diagonal matrices. 

\vspace{0.8cm} 
\noindent
{\large Appendix B}
\renewcommand{\theequation}{B.\arabic{equation}}
\appendix 
\setcounter{equation}{0}
\vspace{0.3cm}
\noindent

 In section 4, we studied the
correspondence between supermembrane and  matrix string 
on the Kaluza-Klein Melvin background.
We started with a light-cone supermembrane action
in flat space-time in 11 dimensions and rewrote it in terms of the
new coordinates which are single-valued along the compactified 
direction.
As a result of the coordinate transformation, 
the action 
in terms of the new coordinates describes 
the supermembrane in a curved background.
In this appendix, as a simple consistency check,
we confirm that the leading part of the expansion of
the curvature ($q^1$-part) of
the resulting membrane action agrees with the 
linearized interaction between supermembrane and backgrounds
derived in \cite{dnp}.
Actually, this problem might sound 
trivial.  However, it is not necessarily so in the presence of 
fermions and 
we decided to include it here for the sake 
of avoiding possible confusion, 
since to relate these two we have to perform a local Lorentz 
transformation for spinors appropriately.  
In the usual treatment of linearized approximation 
such as \cite{dnp}, 
it is not clear which local Lorentz frame is 
used for spinor fields.

The 11-dimensional background which we consider is 
\EQ
ds_{11}^2=-dt^2+dx_1^2+\cdots +dx_6^2 +dr^2+r^2(d\varphi+qdy)^2
+dy^2+dx_{10}^2 .
\label{bg11d}
\EN
This in the linearized approximation $q\ll 1$ 
corresponds to the 
flat 10 dimensional space with constant dilaton 
and a nontrivial magnetic vector field $A_{\varphi}=qr^2$. 
The action at the linear order in $q$ is given as
\EQA
A^{q^1}&=&q\int d\tau d\sigma d\rho\, \epsilon^{mn}
[-D_0YD_0X^{m} X^{n} + \{X^{i},Y\}\{X^{i},X^{m}\}X^{n}\nonumber\\
&&+\{X^{p},Y\}\{X^{p},X^{m}\}X^{n}
-i\psi^T \Gamma_{m} X^{n}\{Y,\psi\}
\nonumber\\
&&-{i\over 4}\psi^T \Gamma_{mn}\psi D_0 Y
-{i\over 4}\psi^T \Gamma_{i}\Gamma_{mn}\psi 
\{ X^{i}, Y \}]
\label{aq1app}
\EQN

Now, in ref \cite{dnp}, `vertex operators for supermembrane'
are derived from the consistency with
supersymmetry, gauge symmetry and with vertex operators
for string or particle upon reduction.
For example, 
the vertex operator for graviton with transverse polarization 
is 
given as
\EQA
V_h&=&h_{ab}[ D_0 X^a D_0 X^b -\{X^a,X^c\}\{X^b,X^c\}
+i\psi\Gamma^a\{X^b,\psi\}
+{1\over 2}D_0X^a \psi^T \Gamma^{bc}\psi k_c 
\nonumber\\
&&+{1\over 2}\{X^a, X^c\}\psi^T\Gamma^{bcd}\psi k_d 
+{1\over 8}\psi^T\Gamma^{ac}\psi\psi^T\Gamma^{bd}\psi k_ck_d ]
\e^{-ik^{a^\prime} X^{a^\prime}},
\label{vh}
\EQN
where we have 
flipped the signs of the coefficients of the fermion 
bilinears from the ones in \cite{dnp} because of the 
difference of our convention. 

Using (\ref{vh}) and the linearization of the 
the background (\ref{bg11d}) 
\[
h_{m9}=-q\epsilon^{mn}x^n=-iq\epsilon^{mn}{\partial\over \partial k^n},
\]
the coupling to the background at $O(q^1)$ is obtained as
\EQA
A_{int}^{(q^1)}&=&q\int d\tau d\sigma d\rho \;\epsilon^{mn}[-D_0YD_0X^m X^n 
+\{X^i, Y\}\{X^i,X^m\}X^n +\{X^p, Y\}\{X^p,X^m\}X^n\nonumber\\
&&-{i\over 2}\psi^T\Gamma^m\{Y,\psi\}X^n
-{i\over 2}\psi^T\Gamma^9\{X^m,\psi\}X^n
-{i\over 4}D_0X^m\psi^T\Gamma^{9n}\psi 
-{i\over 4}D_0 Y \psi^T\Gamma^{mn}\psi\nonumber\\
&&-{i\over 4}\{X^m, X^i\}\psi^T \Gamma^{9in}\psi
-{i\over 4}\{X^m, X^p\}\psi^T \Gamma^{9pn}\psi
-{i\over 4}\{Y, X^i\}\psi^T \Gamma^{min}\psi] .
\label{aint}
\EQN
The purely bosonic part agrees with our result as it should. 
It turns out that the fermion part 
 (\ref{aq1app}) is related to that of (\ref{aint})
by a redefinition of the fermionic fields by
\EQ
\psi\rightarrow \e^{{q\over 4}\epsilon^{mn}X^m\Gamma_{n9}}\psi
=\psi+{q\over 4}\epsilon^{mn}
X^m\Gamma_{n9}\psi+O(q^2) .
\label{rotation}
\EN
This is an allowed transformation, which does not 
change the fermion boundary condition, 
and corresponds to a transformation of local Lorentz 
frame for spinors. In curved space-time, there is always 
such an ambiguity.  In fact, curved space 
action must  be formulated such that it is invariant under the change of local 
Lorentz frame. However, the discussion of vertex operators 
in ref. \cite{dnp}  tacitly assumed a special gauge 
choice for this gauge degrees of freedom to 
ensure manifest world-volume supersymmetry. 
After the rotation (\ref{rotation}), the action acquires additional contribution 
${\tilde{A}}^{q^1}$ to the $q^1$ part
\EQA
{\tilde{A}}^{q^1}&=& q \int d\tau d\sigma d\rho \; \epsilon^{mn}
[-{i\over 4}D_0X^m\psi^T\Gamma^{9n}\psi
-{i\over 4}\{X^m, X^i\}\psi^T \Gamma^{9in}\psi\nonumber\\
&&-{i\over 4}\{X^m, X^p\}\psi^T \Gamma^{9pn}\psi
+{i\over 2}\psi^T X^m\Gamma^9 \{X^n,\psi\}
-{i\over 2}\psi^T X^m\Gamma^n \{Y,\psi\}] .
\EQN
The sum $A^{q^1}+{\tilde{A}}^{q^1}$ is precisely the linearized
coupling obtained from the vertex operator for membranes (\ref{aint}).
 
By making conversion to the matrix-string 
theory following our prescription, 
this should be equivalent with the result we obtain by using the approach 
of ref. \cite{scia}, which is based on the 
standard duality arguments and 
only analyzed, though, the 
zero-th moments of external fields.

\end{document}